\documentclass[11pt,cite,epsfig,psfrag]{article}
\usepackage[utf8]{inputenc}
\usepackage{placeins}
\usepackage{geometry}
 \usepackage{graphicx}
 \usepackage{subfig}
  \usepackage{amsmath} 
  \graphicspath{{images_massive/}}
  \usepackage[usenames, dvipsnames]{color}
\usepackage{wrapfig}
\usepackage{boxedminipage}

\usepackage{fullpage}
\usepackage{amsmath}
\usepackage{amssymb}
\usepackage{graphicx}
\usepackage{mathrsfs}
\usepackage{wrapfig}
\usepackage{boxedminipage}
\usepackage{epsfig}
\usepackage{setspace}
\usepackage{float}
\usepackage{hyperref}
\usepackage{enumerate}
\usepackage{cite}
\usepackage[font=footnotesize,labelfont=rm]{caption}

\usepackage[numbers,sort&compress]{natbib}

\def\be{\begin{equation}}
\def\ee{\end{equation}}
\def\bea{\begin{eqnarray}}
\def\eea{\end{eqnarray}}

\numberwithin{equation}{section}

 \newcommand{\RN}[1]{%
   \textup{\uppercase\expandafter{\romannumeral#1}}%
 }
\begin{document}

\thispagestyle{empty}

\vskip 2cm

\begin{center}
{\Large \bf Critical Heat Engines in Massive Gravity}
\end{center}

\vskip .2cm

\vskip 1.2cm

\centerline{ \bf   Pavan Kumar Yerra \footnote{pk11@iitbbs.ac.in} and Chandrasekhar Bhamidipati\footnote{chandrasekhar@iitbbs.ac.in}
}

\vskip 7mm 
\begin{center}{ School of Basic Sciences\\ 
Indian Institute of Technology Bhubaneswar \\ Bhubaneswar, Odisha, 752050, India}
\end{center}

\vskip 1.2cm
\vskip 1.2cm
\centerline{\bf Abstract}
\noindent
With in the extended thermodynamics, we study the efficiency $\eta_k$ of critical heat engines for charged black holes in massive gravity for spherical ($k=1$), flat ($k=0$) and hyperbolic ($k=-1$) topologies. Although, $\eta_k$ is in general higher (lower) for hyperbolic (spherical) topology, we show that this order can be reversed in critical heat engines with efficiency higher for spherical topology, following in particular the order: $ \eta_{\rm -1}^{\phantom{-1}} < \eta_{\rm 0}^{\phantom{0}} < \eta_{\rm +1}^{\phantom{+1}}$. Furthermore, the study of the near horizon region of the critical hole shows that,  apart from the known $q\rightarrow \infty $ condition, additional scalings of massive gravity parameters, based on the topology of the geometry are required, to reveal the presence of a fully decoupled Rindler space-time with vanishing cosmological constant. 

\newpage
\setcounter{footnote}{0}
\noindent

\baselineskip 15pt

\section{Introduction}

Investigations of the critical region of black holes in AdS~\cite{Chamblin:1999tk,PhysRevD.60.104026,Caldarelli:1999xj}, particularly, in the context of extended phase space approach~\cite{Kastor:2009wy,Dolan:2011xt,Kubiznak:2012wp,Gunasekaran:2012dq} has been an exciting area of research. In the neighborhood of a second order phase transition, the thermodynamic quantities of charged black holes in AdS, turn out to show scaling behavior, with respect to charge $q$, i.e., Entropy $S\sim q^2$, Pressure $p\sim q^{-2}$, and Temperature $T\sim q^{-1}$.  More intriguingly, the  black hole geometry turns out to be a fully decoupled Rindler space-time in the double scaling limit of: nearing the horizon while also taking the charge to be large~\cite{Johnson:2017hxu,Johnson:2017asf}. In the past, the emergence of such decoupled space-times in the near horizon limit of charged black holes leading to extremal black holes have generated enormous interest, such as, microscopic counting of black hole entropy, among other issues involving AdS/CFT duality. The appearance of fully decoupled Rindler geometries in the context of critical charged black holes in AdS is quite non-trivial and should lead to novel results from CFT point of view in this scenario. It is important to know if the above mentioned features are generic to black holes in AdS or not.  The scaling behavior of thermodynamic quantities mentioned above, of course varies depending on the critical behavior of black holes in question and can also be mildly broken if one of the thermodynamic quantities depends on additional parameters (as will be the case in the model considered in this paper) and the near horizon limit in this case needs to be reexamined. We will see that the setting of considering black holes in massive gravity theories gives interesting results on the aforementioned issues, which are motivated at the end of this section. \\

\noindent
Novel studies of the critical region were in fact possible only due to active developments in extended thermodynamic description of charged black holes in AdS, which reveal a phase structure consisting of line of first order phase
transitions terminating in a second order transition point~\cite{Chamblin:1999tk,Kubiznak:2012wp,Gunasekaran:2012dq,Cai:2013qga,Kubiznak:2016qmn,Pedraza:2018eey}. Here, apart from continued research on Hawking-Page transition in the bulk (holographically dual to confinement-deconfinement transition in gauge theories), van der Waals transition has aso attracted wide spread attention, with a  holographic interpretation being actively pursued. For instance, in~\cite{Johnson:2014yja} (see also~\cite{Kastor:2009wy,Dolan:2013dga}), the consequence of considering a variable cosmological constant, is thought to be a tour in the space of  dual field theories (labeled by N, the number of colors in the gauge theory). Varying the cosmological constant in the bulk may correspond to perturbing the dual CFT, triggering a field theory renormalization group flow. This flow in the bulk is expected to be captured by Holographic heat engines with black holes playing the role of working substances~\cite{Johnson:2014yja}. Specifically, for hyperbolic space-times, the efficiency of heat engines may have a nontrivial connection with central charges and degrees of freedom of dual CFT~\cite{Johnson:2018amj}. We should mention, that this is currently a very interesting topic with various aspects being studied both from the gravity as well as the dual gauge theory side~\cite{Dolan:2011xt,Dolan:2010ha,Kastor:2009wy,Caldarelli:1999xj,Sinamuli:2017rhp,Karch:2015rpa,Kubiznak:2014zwa,Johnson:2014yja,Johnson:2015ekr,Johnson:2015fva,Belhaj:2015hha,Bhamidipati:2016gel,Chakraborty:2016ssb,Hennigar:2017apu,Johnson:2017ood,Setare2015,Caceres2015,Mo2017,Liu:2017baz,Wei:2016hkm,Sadeghi:2015ksa,Zhang:2016wek,Kubiznak:2016qmn,Sadeghi:2016xal,Ghaffarnejad:2018gtj}.\\

\noindent
There is another interesting context in which the critical region of black holes in AdS plays a central role, namely, for improving the efficiency of heat engines.
 It has been shown recently, that the efficiency of heat engines when the black hole is on the verge of a second order phase transition, leads to the interesting possibility of reaching Carnot efficiency. With this hope, following the works of Johnson~\cite{Johnson:2017hxu}, efficiency of critical black hole heat engines has been computed for several systems, involving Gauss-Bonnet and non-linearly charged black holes. In all the cases, it was noted that the engine efficiency reaches the Carnot efficiency, however only in the limit that the engine runs for an infinite time\footnote{see e.g.~\cite{PhysRevLett.106.230602,PhysRevLett.111.050601,PhysRevLett.115.090601,PhysRevX.5.031019,PhysRevLett.114.050601,power_of_a_critical_heat,Koning2016,PhysRevLett.117.190601,eff_vs_speed}, for ongoing work on approaching Carnot efficiency in statistical mechanics literature.}, as certain parameters of the engine are taken to be large~\cite{Johnson:2017hxu,Bhamidipati:2017nau,Yerra:2019fxh}. Building on the earlier works, in this paper, we study the efficiency of critical heat engines with black holes in massive gravity as working substances. In the next paragraph, we present a general motivation for studying massive gravity theories and following that, we give a specific reasons for choosing this system for studying critical heat engines.\\

\noindent
Broad motivations for studying massive gravity theories follow.
Einstein's General relativity has met with lot of success, with important predictions having received experimental 
 confirmation, more recently in agreement with recent
observational data of LIGO collaboration~\cite{LIGO2017,deRham2014Review} on gravitational waves. However, there are also phenomena which, such as, accelerated expansion of the universe and the cosmological constant problem, to name a few, which warrant extensions of the Einstein's theory. In this context, an important extension involves massive graviton theories, motivated by hierarchy problems and their usefulness in quantum gravity~\cite{MassiveIb,MassiveIc}, which binds well with recent data~\cite{Abbott}, putting lower limits on the mass of gravitons. Massive gravity theories have long history, starting from the models introduced by Fierz and Paullo in 1939~\cite{Fierz1939}, which underwent several modifications and inclusion of novel ideas, such as, New massive gravites\cite{BDghost,Newmasssive,dRGTI,dRGTII}, which has been actively studied in current literature \cite{NewM1,NewM2,NewM3,NewM4,NewM5,HassanI,HassanII}. Black hole solutions, their thermodynamical properties~\cite{BHMassiveI,BHMassiveII,BHMassiveIII,BHMassiveIV} and applications in cosmology/astrophysics with motivations to see deviations from Einstein's General relativity are being actively pursued too~\cite{Katsuragawa,Saridakis,YFCai,Leon,Hinterbichler,Fasiello,Bamba}. One class of massive gravity theories with possible applications to holographic duality was considered in~\cite{Vegh}, involving the use of a singular metric, showing that the massive
gravity might be stable and free of ghosts \cite{HZhang}, including the presence of black hole 
solutions\cite{PVMassI,PVMassII,PVMassIII,PVMassIV,PVMassV}. Massive gravity theories are expected to play important role in solving problems discussed above in Einstein's gravity~\cite{Gumrukcuoglu,Gratia,Kobayash,DeffayetI,DeffayetII,DvaliI,
DvaliII,Will,Mohseni,GumrukcuogluII,NeutronMass,Ruffini}, such as, the ability to explain the
current observations related to dark matter
\cite{Schmidt-May2016DarkMatter} and also the accelerating
expansion of universe without requiring any dark energy component
\cite{MassiveCosmology2013,MassiveCosmology2015}. Attempts to embed massive gravities in string theory are being pursued too~\cite{MGinString2018}. More importantly, Van der Waals type liquid gas phase transitions in the extended phase space have been shown to exist and studied in a number of works~\cite{Cai2015,PVMassV,PVMassIV,Alberte,Zhou,Dehyadegari,Magmass}. \\

\noindent
Now we describe specific motivation for studying critical heat engines in massive gravity theories. First, the massive gravity theories considered in this paper can be regarded as the minimal modification of general relativity which takes into account a massive graviton. The effect of introducing graviton mass is phenomenal as it leads to the existence of van der Waals phase transitions for non-spherical topologies, which are forbidden in Einstein as well as higher curvature Lovelock gravity. It is then interesting to see how the efficiency of heat engines varies with topology of the black hole. Preliminary studies in this direction were undertaken in~\cite{Hendi:2017bys}, showing in particular that black holes with hyperbolic horizons as heat engines turn out to have maximum efficiency, followed by flat and spherical horizon cases. Our aim in this work is to check whether the above dependence of efficiency on horizon topology, continues to hold when the heat engine is operated close to the critical point in thermodynamic phase space. The reason for checking this, is that efficiency of heat engines is very sensitive to the scheme  chosen for computations\footnote{There are actually various schemes possible, obtained by choosing a thermodynamic cycle in which certain parameters are held fixed and others varied. For instance, one possible scheme involves picking operating pressures and temperatures, with volume being left  unfixed (to be determined from equation of state) and so on.} and can show an increase in one scheme and a decrease in another~\cite{Johnson:2015ekr}; and also show certain universal features when the engine runs close to criticality in the large charge limit~\cite{Johnson:2017hxu,Johnson:2017asf}. We thus, analyze properties of heat engines in massive gravity at the critical point and compare the efficiencies for horizons of various topologies. Interestingly, we find that efficiencies are highest for spherical topologies, followed by flat and hyperbolic horizon cases, a result, quite opposite to the situation when the engine runs far from criticality~\cite{Hendi:2017bys}. A second motivation is that, the study of the near horizon limit of all charged or neutral critical black holes in AdS, in the large charge limit, thus far has shown the emergence of a fully decoupled Rindler geometry, which is because of a perfect scaling of all thermodynamic quantities with charge $q$ at the critical point (as mentioned in the first paragraph of this section). However, as we see below, one of the effects of massive graviton is to spoil the perfect of scaling of temperature with respect to charge (see eqn.(\ref{eq:critical pt}) to be discussed later). We show that the appearance of a fully decoupled Rindler space-time may or may not appear in the near horizon limit, depending on the values chosen for massive gravity parameters.\\

\noindent
Rest of the paper is organized as follows. In section-(\ref{2}), we give brief details of charged black holes in AdS in massive gravity theories and collect results on various thermodynamic quantities, including the equation of state and the PV critical behavior. In section-(\ref{3}), we set up the computation of efficiency of heat engines at the critical point and bring out the role played by the massive gravity parameters. The results on efficiency at the critical point are compared for various topologies. We also analyze the critical region of black hole and present the conditions under which a fully decoupled Rindler space-time appears in the near horizon limit. We summarize our findings in section-(\ref{conclusions}).

\section{Charged  Black Holes in Massive Gravity} \label{2}
Consider the action for 4-dimensional Einstein-Maxwell theory with a negative cosmological constant 
$\Lambda$ in massive gravity as~\cite{Cai2015,PVMassIV,Hendi:2017bys}:
\begin{equation}
I=-\frac{1}{16\pi }\int d^{4}x\sqrt{-g}\left( \mathcal{R}-2\Lambda
-\mathcal{F}+m^{2}\sum_{i}^{4}c_{i}\mathcal{U}_{i}(g,f)\right) ,
\end{equation}
where  $\mathcal{R}$ is the Ricci scalar, $\mathcal{F}=F_{\mu \nu
}F^{\mu \nu }$ is the Maxwell invariant,  $F_{\mu \nu }=\partial _{\mu
}A_{\nu }-\partial _{\nu }A_{\mu }$ is the electromagnetic field  tensor with gauge
potential $A_{\mu }$, and  $m$ is mass parameter for graviton.  The $c_{i}$'s are   constants, $f$ is a reference metric  and the $\mathcal{U}_{i}$'s
are symmetric polynomials of the eigenvalues of the $4\times 4$ matrix $ \mathcal{K}_{\nu }^{\mu }=\sqrt{g^{\mu \alpha }f_{\alpha \nu }}$, which can be written as
\begin{eqnarray}
	\mathcal{U}_{1}&=&\left[ \mathcal{K}\right], \nonumber \\ \mathcal{%
		U}_{2} &= & \left[ \mathcal{K}\right] ^{2}-\left[ \mathcal{K}^{2}\right] , \nonumber \\ 
	\mathcal{U}_{3}&=&\left[ \mathcal{K}\right] ^{3}-3\left[ \mathcal{K}\right] %
	\left[ \mathcal{K}^{2}\right] +2\left[ \mathcal{K}^{3}\right], \nonumber \\ 
	\mathcal{U}_{4}&=&\left[ \mathcal{K}\right] ^{4}-6\left[ \mathcal{K}^{2}%
	\right] \left[ \mathcal{K}\right] ^{2}+8\left[ \mathcal{K}^{3}\right] \left[
	\mathcal{K}\right] +3\left[ \mathcal{K}^{2}\right] ^{2}-6\left[ \mathcal{K}%
	^{4}\right].
\end{eqnarray}
The above action admits the static topological black hole solution with the metric~\cite{Cai2015,PVMassIV,Hendi:2017bys}:
\begin{equation}
ds^{2}=-Y(r) dt^{2}+\frac{dr^{2}}{Y(r) } + r^{2}h_{ij}dx_{i}dx_{j} \ ,
  \label{eq:metric}
\end{equation}
and together with a reference metric $f_{\mu \nu}$:
\begin{equation} f_{\mu \nu
}=\text{diag}(0, 0, c_0^{2} h_{ij}) \ ,
\end{equation}
where $c_0$ is a positive constant,  $i,j=1,2$ and $h_{ij}dx_{i}dx_{j}$ is a spatial metric of
constant curvature $2k$ with volume $4\pi$. Here, one can take $ k $ = +1, 0, or -1, for  a spherical, Ricci flat, or hyperbolic topology of the black hole horizon, respectively.
Using the reference metric $f_{\mu \nu}$, the $\mathcal{U}_{i}$'s are read as~\cite{Cai2015,PVMassIV,Mo:2017nes} 
\begin{equation}
\mathcal{U}_{1}= \frac{2c_0}{r}, \quad \mathcal{%
	U}_{2} =  \frac{2c_0^2}{r^2} , \quad
\mathcal{U}_{3}= 0 , \quad
\mathcal{U}_{4}= 0 \ ,
\end{equation}
where one can set  $c_3 =c_4 = 0$, since $\mathcal{U}_{3}= \mathcal{U}_{4}= 0$. The metric function $Y(r)$, using the guage potential ansatz $A_{\mu }=h(r)\delta
_{\mu }^{0}$ where $h(r) = (\frac{q}{r_+} - \frac{q}{r})$~\cite{Dehghani:2019thq},  is given by~\cite{Cai2015,PVMassIV,Hendi:2017bys}:
\begin{equation}
Y(r) =k-\frac{m_{0}}{r}-\frac{\Lambda r^{2}}{3}+\frac{q^{2}}{ r^{2}}+m^{2}(\frac{c_0c_{1}}{2}r+c_0^{2}c_{2}) \ ,
 \label{Y(r)}
\end{equation}
where the integration constants  $m_0$ and $q$ correspond to the mass $M$ and the electric charge $Q$ of the black hole, respectively.
The solution~\eqref{Y(r)}, is asymptotically AdS and in the absence of graviton mass $(m =0)$, it reduces to the AdS Reissner-Nordstrom black hole~\cite{Cai2015,PVMassIV}. We note that the choice of the reference metric makes the graviton mass terms to have a Lorentz-breaking property~\cite{Vegh:2013sk}. 

\vskip 0.5cm
\noindent
The  horizon radius $r_+$ of the black hole is  the largest positive root of $Y(r_+)=0$, in terms of which the temperature $T$, mass $M$, entropy $S$,  charge $Q$, and the electric potential $\Phi$ of the black hole can be expressed as~\cite{Cai2015,PVMassIV}:

\begin{eqnarray}
T &=&\frac{k}{4\pi r_{+}}-\frac{r_{+}\Lambda }{4\pi
}-\frac{q^{2}}{4\pi r_{+}^{3}}  +\frac{m^{2}}{4\pi r_{+}}\left(
c_0c_{1}r_{+}+c_{2}c_0^{2}\right) \ ,
\label{eq:Temp} \\
 M &=&\frac{m_{0}}{2 } = \frac{r_+}{2} \left(k -\frac{\Lambda}{3}r_+^2 + \frac{q^2}{r_+^2} + m^2(\frac{c_0 c_1}{2}r_+ + c_0^2c_2)\right)\ ,  \label{eq:Mass} \\
S &=&\pi r_{+}^{2} \ ,\\
Q &=& q \ , \\
\Phi &=& A_{\mu }\chi ^{\mu }\left\vert _{r\rightarrow \infty }\right. -A_{\mu
}\chi ^{\mu }\left\vert _{r\rightarrow r_{+}}\right. =\frac{q}{r_{+}} \ .
\end{eqnarray}
\noindent
In the extended phase space, we define the pressure from assuming a dynamical cosmological constant\footnote{A dynamical cosmological constant scenario can occur in various methods in a situation where the gravity theory under study is embedded in a larger set up, involving other matter fields, possibly of origin in string theory. When the high energy theory has several dynamical scalar fields with a suitable potential, the fixed points of this potential give non-zero vev's to the scalars, which can be treated as a cosmological constant, see e.g.,~\cite{Aharony:1999ti}. We do not pursue these aspects here.}, using $p=-\frac{\Lambda}{8\pi}$, and its conjugate quantity is the thermodynamic volume $V$. Then one should identify the mass $M$ of the black hole as the enthalpy $H$~\cite{Kastor:2009wy}, which satisfies the first law of 
black hole thermodynamics~\cite{Cai2015,PVMassIV}:

\begin{equation}
dM=TdS+\Phi dQ+Vdp+\mathcal{C}_{1}dc_{1} \ ,  \label{1stlaw}
\end{equation}
where 
\begin{eqnarray}
V &=&\left( \frac{\partial M}{\partial p}\right) _{S,Q,c_{1}}=\frac{4\pi}{3} r_+^3 \ ,  \label{Vol} \\
\mathcal{C}_{1} &=&\left( \frac{\partial M}{\partial c_{1}}\right)
_{S,Q,p}=\frac{c_0m^{2}r_{+}^{2}}{4 } \ .
\label{C1} 
\end{eqnarray}
Now, using $p=-\frac{\Lambda}{8\pi}$, in equation~\eqref{eq:Temp} and equation~\eqref{eq:Mass}, we obtain the expressions for equation of state $p(V, T)$ and enthalpy $H(S,p)$ as: 
\begin{eqnarray}
p & = & \frac{1}{8\pi}\Bigg\{\frac{(4\pi T-m^2c_0c_1)}{(\frac{3V}{4\pi})^{\frac{1}{3}}}-\frac{(k+m^2c_2c_0^2)}{(\frac{3V}{4\pi})^{\frac{2}{3}}}+\frac{q^2}{(\frac{3V}{4\pi})^{\frac{4}{3}}}\Bigg\} \ , \label{eq of st} \\
H &\equiv& M = \frac{1}{6\sqrt{\pi S}} \Bigg\{  8pS^2 +3S(k+m^2c_2c_0^2) + 3\pi q^2 + \frac{3m^2c_0 c_1}{2}\sqrt{\frac{S^3}{\pi}} \Bigg\} \ . \label{eq: enthalpy}
\end{eqnarray}
The presence of massive graviton could admit the critical behavior for the black holes with topology flat  $(k=0)$ and  hyperbolic $(k=-1)$ as well, unlike the case of massless graviton, where only the black holes with spherical topology $(k=+1)$ can exhibit the critical behavior~\cite{PVMassIV,Kubiznak:2012wp}.
\begin{figure}[h]
	\begin{center}
		\centering
		\includegraphics[width=4in]{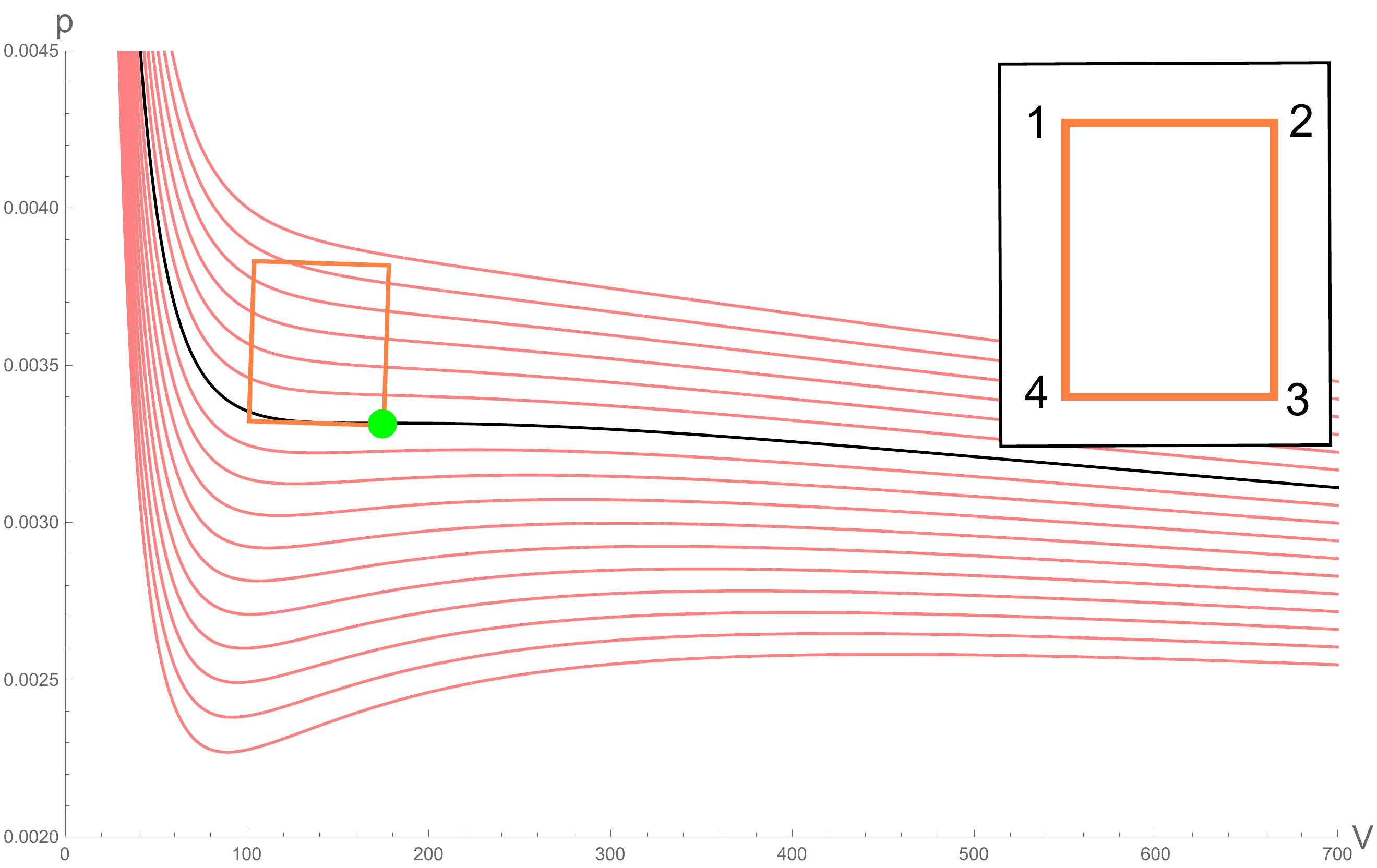}  
		
		\caption{ Sample isotherms in $p-V$ plane for the equation of state~\eqref{eq of st}. The central (black) isotherm is for critical temperature $T_{cr}$; the temperature of the isotherms decreases from top to bottom  and the critical point is highlighted with green colored dot where  the corner 3 of the engine cycle is  placed (see the inset for  labeling of the cycle). Here, the parameters $k =-1, q=2, m=c_0=1, c_1=0.01, \text{and} \, c_2=3$, are used and a similar phase structure exists for other topologies.}   \label{fig:sample isotherms}
		
	\end{center}
\end{figure}
The equation of state~\eqref{eq of st}, facilitates the study of critical behavior of  topological black holes on plotting different isotherms as shown in figure~(\ref{fig:sample isotherms}). For fixed parameters $(k, q, m, c_0, c_1,  c_2)$, there  exists a critical temperature $T_{cr}$, corresponding to a critical isotherm. The isotherms above the critical isotherm, have the   behaviour  of an ideal gas and indicate the existence of unique phase of the black holes, while,
those below the critical isotherm, show oscillatory behavior that indicate the small and large black hole phase. These small and large black holes undergo a first order phase transition that terminates at the critical point, from where the phase transition is of second order.      
This phase structure of the topological black holes in massive gravity is a reminiscent of the liquid/gas phase transition of van der Waals fluid~\cite{PVMassIV,ston201john4thermodynamic}.
 The critical point can be obtained from the condition of stationary point of inflection $(i.e.,  \partial p/\partial V=\partial^2 p /\partial V^2=0  )$, given by~\cite{PVMassIV}:  
\begin{equation}
	\label{eq:critical pt}
	\quad p_{\rm cr}=\frac{\epsilon^2}{96\pi q^2 }\ , \quad V_{\rm cr}=\frac{8 \sqrt{6}\pi q^3}{\epsilon^{\frac{3}{2}}} \ ,   \quad T_{\rm cr}= \frac{\epsilon^{\frac{3}{2}}}{ 3\sqrt{6} \pi q} + \frac{m^2c_1 c_0}{4\pi} \ , 
\end{equation}
where $\epsilon = (k +m^2c_2 c_0^2) > 0$. We also note that $V_{\rm cr}$ in eqn. (\ref{eq:critical pt}) can be calculated in terms of the critical radius $r_{cr}$. $r_{cr}$ can be found by expressing the equation of state in eqn.(\ref{eq of st}) in terms of the horizon radius $r_+$ and finding the points of inflection, which yield $r_{cr}=\sqrt{\frac{6}{\epsilon}}q $. Substituting $r_{cr}$ in eqn.(\ref{Vol}) gives the critical volume $V_{cr}$.
 Furthermore, the specific heats of the black holes at constant volume $C_V$, and at constant pressure $C_p$ are given by~\cite{Hendi:2017bys,Mo:2017nes}:  
 \begin{equation}
	C_V=0 \ ; \, \, \,   C_p= 2S\Bigg(\frac{8pS^2+S(k+m^2c_0^2 c_2) -\pi q^2 + \frac{m^2c_0 c_1 S^{3/2}}{\sqrt{\pi}}}{8pS^2 -S(k+m^2c_0^2 c_2) + 3\pi q^2 } \Bigg) \ .
\end{equation}
The specific heat $C_p$ can indeed be positive at least in certain regions of parameters space and hence the construction of heat engines can be done.
\vskip 0.5cm

\section{Critical Heat Engines in Massive Gravity} \label{3}
\noindent
With the set up of extended thermodynamics given in last section, one can proceed to define heat engines for extracting mechanical work from heat energy via the $pdV$ term present in the First  Law of extended black hole thermodynamics~\cite{Johnson:2014yja}, where, the working substance is a black hole solution of the massive gravity system satisfying the equation of state given in eqn. (\ref{eq of st}). First step is to define a cycle in thermodynamic state space with input heat flow $Q_H$, output heat flow $Q_C$, and a net output work W, satisfying the relation $Q_H = W + Q_C$. The efficiency of heat engines can then be written in the well known way as $\eta=W/Q_H=1-Q_C/Q_H$. Actual computation of efficiency can be done by evaluating $\int C_p dT$ along the isobars, where~$C_p$ is the specific heat at constant pressure or more efficiently via the exact formula given in~\cite{Johnson:2015ekr,Johnson:2015fva,Johnson:2016pfa}: 
\begin{equation}
\eta = 1- \frac{M_3 - M_4}{M_2 - M_1} \, , \label{eq:efficiency-prototype} 
\end{equation}
which needs to be evaluated at all four corners of the cycle. We define  a rectangle in $p-V$ plane as our engine cycle (which is a natural choice for static black holes with $C_V = 0$ ~\cite{Johnson:2014yja}), and compute its  efficiency  using the exact formula~\eqref{eq:efficiency-prototype}. 

\subsection{Efficiency at Criticality}
It is noted in~\cite{PhysRevLett.114.050601,power_of_a_critical_heat} that, running the engine cycle in  the vicinity of  critical point leads to approaching the Carnot's efficiency with non zero power. This novel feature was also realized in the context of black holes by considering the limit of certain parameters to be large: such as charge~\cite{Johnson:2017hxu} or other couplings of theories under consideration~\cite{Bhamidipati:2017nau,Yerra:2019fxh}. Taking advantage of the existence of critical region, one can put the corner 3 of the cycle at critical point (see fig. \ref{fig:sample isotherms}) and take the boundaries of the cycle\footnote{Other choices are also permissible and equivalent.} in the following way:
\begin{eqnarray}\label{cyclecrit}
p_3 &= &p_4 =p_{\rm cr}, \nonumber \\ 
p_1 &= & p_2  =  3p_{\rm cr}/2,  \nonumber \\
V_2 &=& V_3=V_{\rm cr},  \nonumber \\ \text{and} \, \, \, \,  V_1 &=& V_4= V_{\rm cr}\Big(1-\frac{L}{q\sqrt{\epsilon}}\Big) \ , 
\end{eqnarray}
where  $L$ is a constant with dimensions of charge and $0 <\frac{L}{q\sqrt{\epsilon}} <1$. This set up makes the work done $W$ (which is simply the area of the cycle) to be a constant and independent of charge~\cite{Johnson:2017hxu}, obtained to be:
\begin{equation}\label{eq:work}
W =\frac{L}{4\sqrt{6}} \ .
\end{equation}
In fact, the scalings in eqn. (\ref{cyclecrit}) have been chosen in such a way that $W$ is even independent of topology, which is quite useful as we will see later~\footnote{ More general choices and also various scalings of the engine together with their limitations are  elaborated in Appendix A. Furthermore, details of the scheme independent behavior of efficiency are pointed out in Appendix B.}. The heat flow on the other hand is seen to be:
\begin{eqnarray}
Q_H &=& M_2 -M_1 \nonumber \\
  &=& \Bigg\{ { \frac{3L}{4\sqrt{6}} + \frac{3c_0 c_1 m^2 q^2}{2 \epsilon}}\bigg[1-\Big(1 - \frac{L}{q\sqrt{\epsilon}}\Big)^{\frac{2}{3}}\bigg]   + \frac{q}{2}\sqrt{\frac{\epsilon}{6}}\bigg[1-\Big(1-\frac{L}{q\sqrt{\epsilon}}\Big)^{-\frac{1}{3}}\bigg] \nonumber \\
   && \,  + \, \,  3q\sqrt{\frac{\epsilon}{6}}\bigg[1-\Big(1-\frac{L}{q\sqrt{\epsilon}}\Big)^{\frac{1}{3}}\bigg]  \Bigg\} \, ,
\label{QH} \end{eqnarray}
and explicitly depends on charge as well as topology.
Engine efficiency is given as $\eta = W/Q_H$,  while the Carnot efficiency $\eta_{\rm C}^{\phantom{C}}$ is calculated from the highest $(T_H =T_2)$ and lowest $(T_C = T_4)$ temperatures  using the equation of state~\eqref{eq of st} as:
\begin{eqnarray}
\label{eq:itac}
\eta_{\rm C}^{\phantom{C}} &=& 1-\frac{T_C}{T_H} \nonumber \\
&=& \frac{\bigg[19 -6\Big(1-\frac{L}{q\sqrt{\epsilon}}\Big)^{\frac{1}{3}} -12\Big(1-\frac{L}{q\sqrt{\epsilon}}\Big)^{-\frac{1}{3}} + 2\Big(1-\frac{L}{q\sqrt{\epsilon}}\Big)^{-1}\bigg]}{\Big[19 +\frac{12\sqrt{6}c_0 c_1 m^2q}{\epsilon^{3/2}}\Big]}
\ .\end{eqnarray}
Now, we can examine the behavior of efficiency in two special cases, i.e., when the  massive coefficient $c_1 = 0$ and $c_1 \neq 0$. We will also study efficiency as a function of graviton mass $m$. Without loss of generality, one can set the constants $c_0$ and $c_2$ to fixed values for the above study.\\

\begin{figure}[h!]
 	{\centering
 		\subfloat[]{\includegraphics[width=2in]{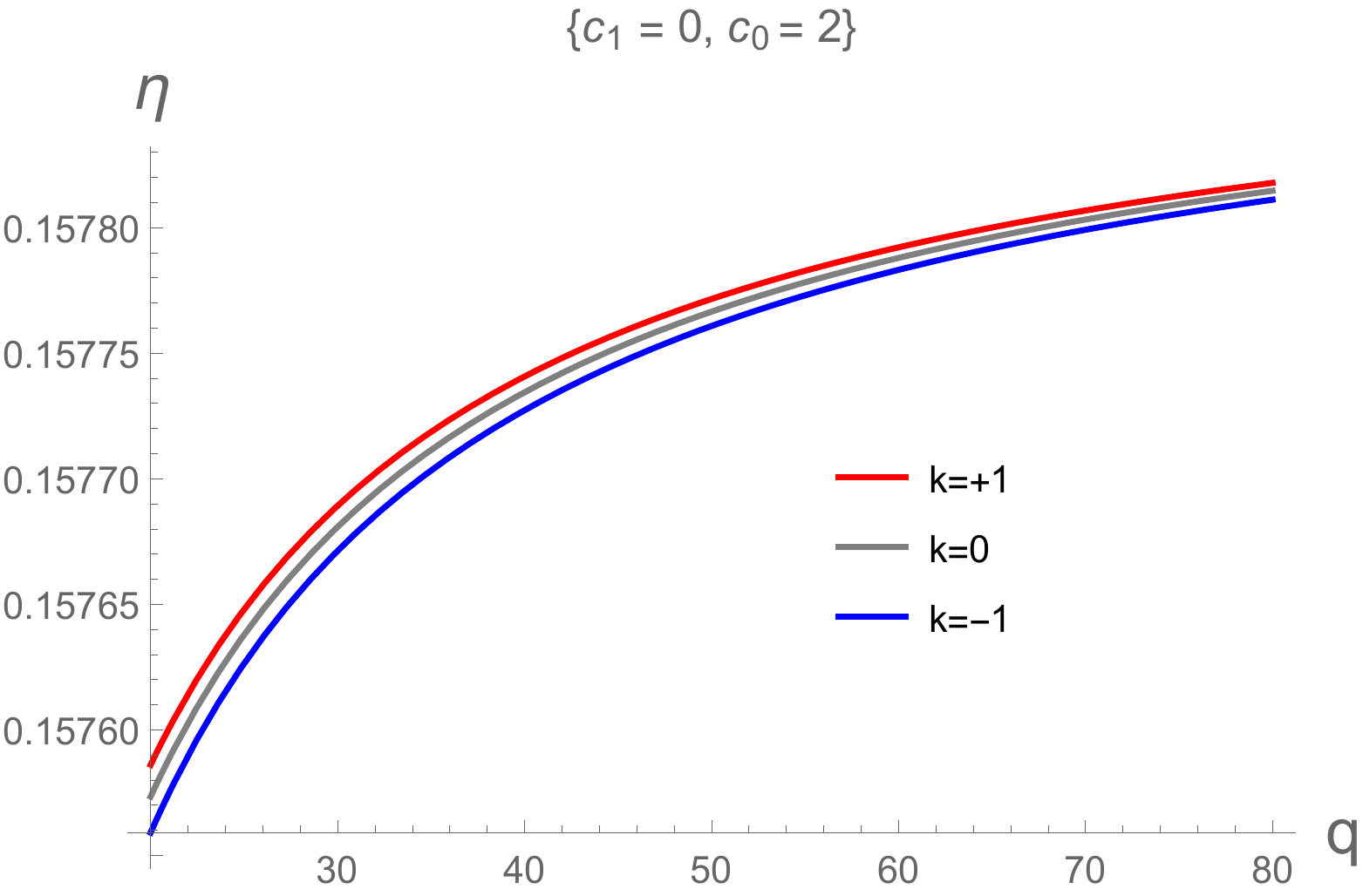}}\hspace{0.15cm}
 		\subfloat[]{\includegraphics[width=2in]{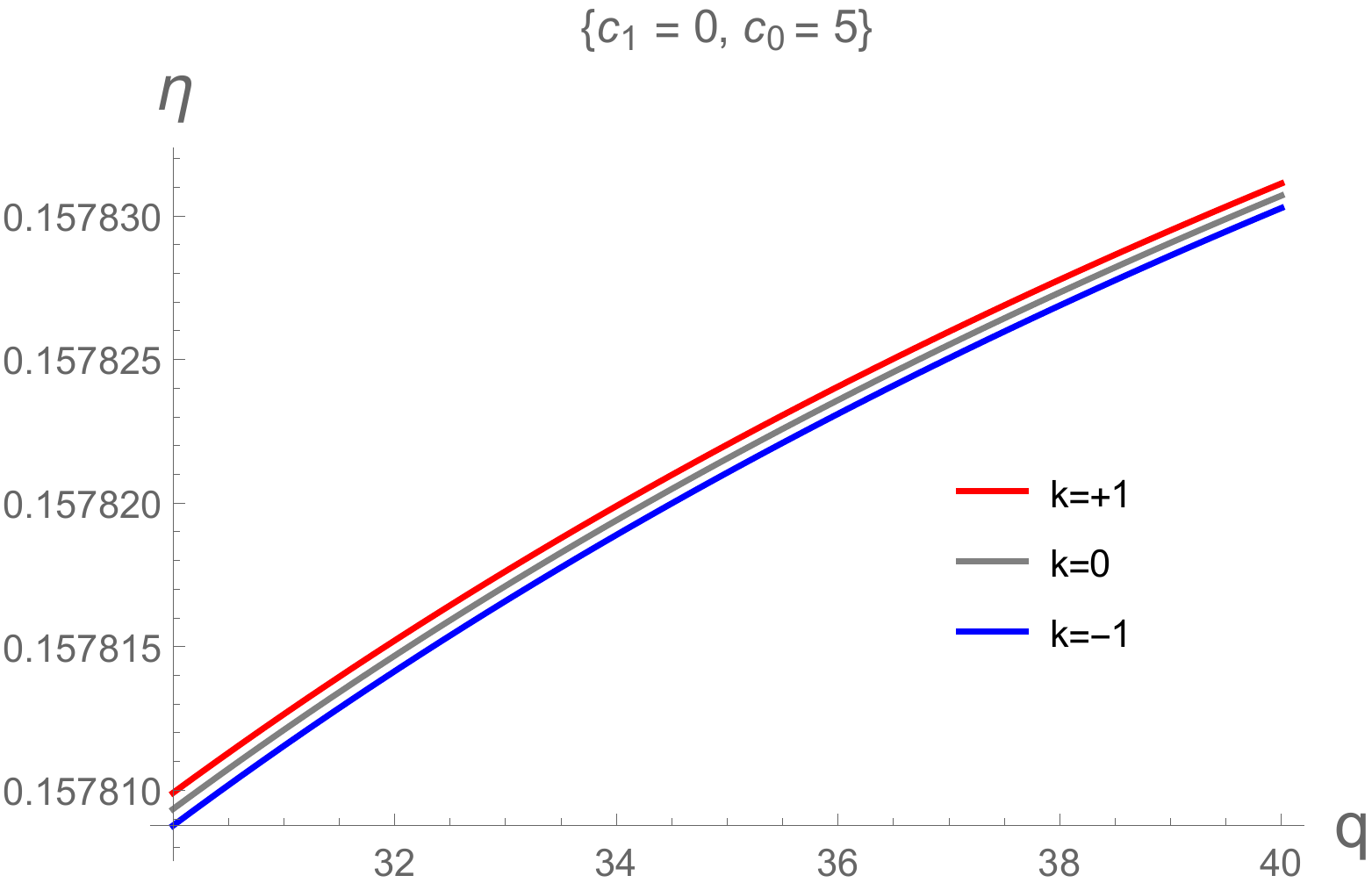}}\hspace{0.15cm}
 		\subfloat[]{\includegraphics[width=2in]{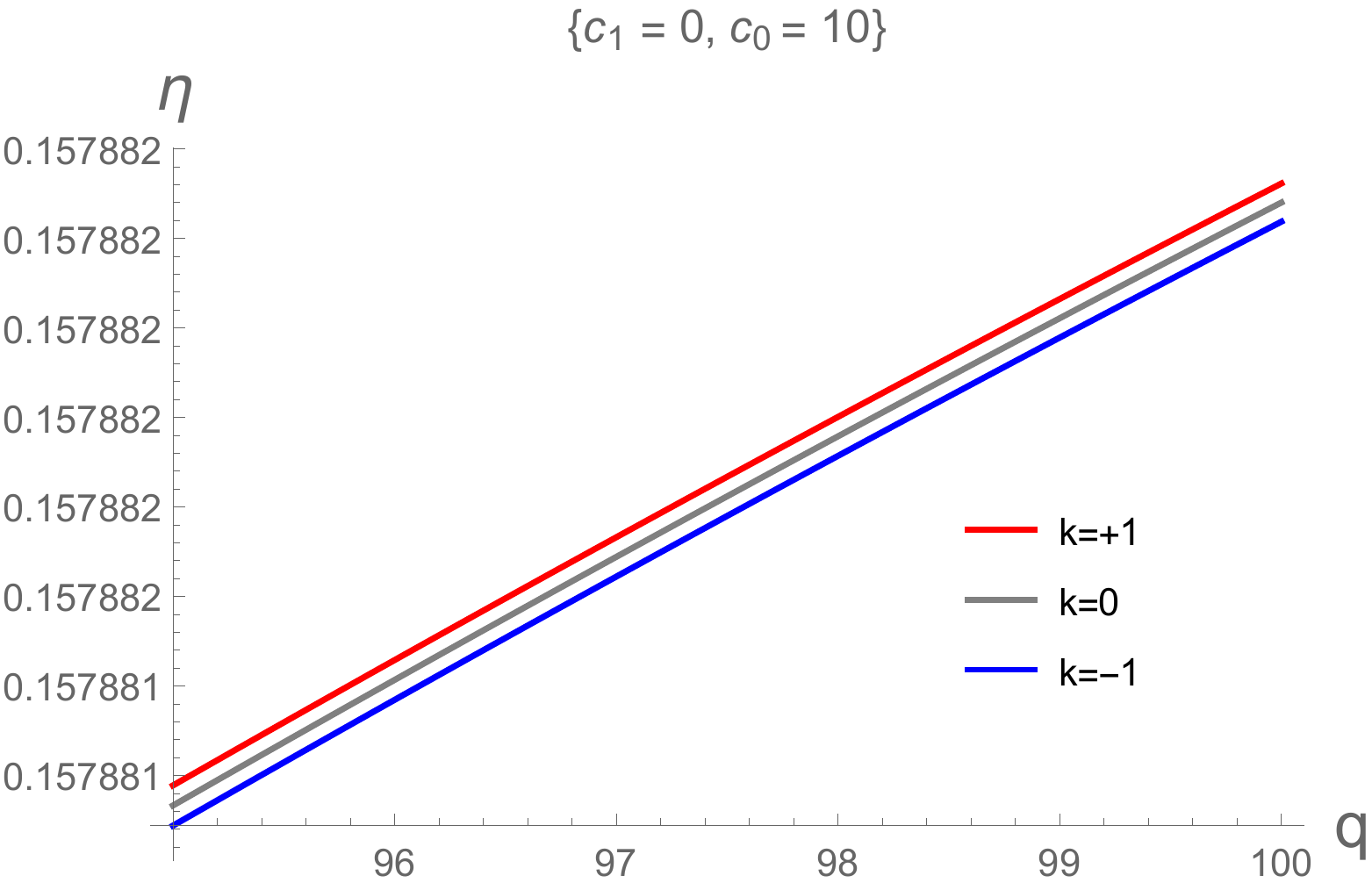}}

 		\caption{\footnotesize In the case of massive coefficient $c_1=0$, the behavior of efficiency $\eta$ with topology $k$ at various $c_0$ (a) $c_0 = 2$, (b) $c_0 = 5$ and (c) $c_0 = 10$. (Here, the parameters $L= m =1$, and  $c_2 =3$, are used.) 
 		}\label{fig:c0_sensitive}  
 	}
 \end{figure}

\begin{figure}[h!]
 	{\centering
 		\subfloat[]{\includegraphics[width=2in]{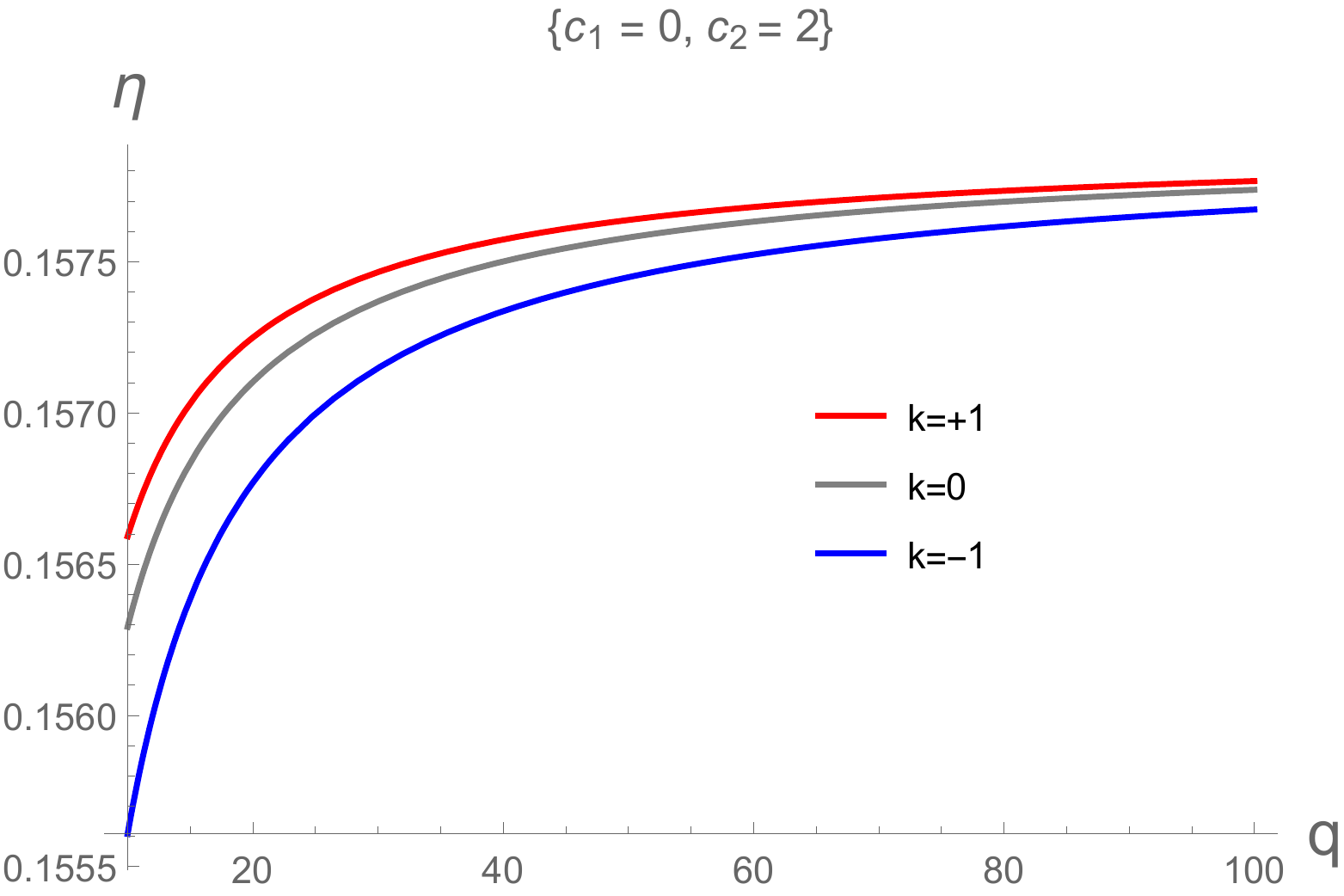}}\hspace{0.15cm}
 		\subfloat[]{\includegraphics[width=2in]{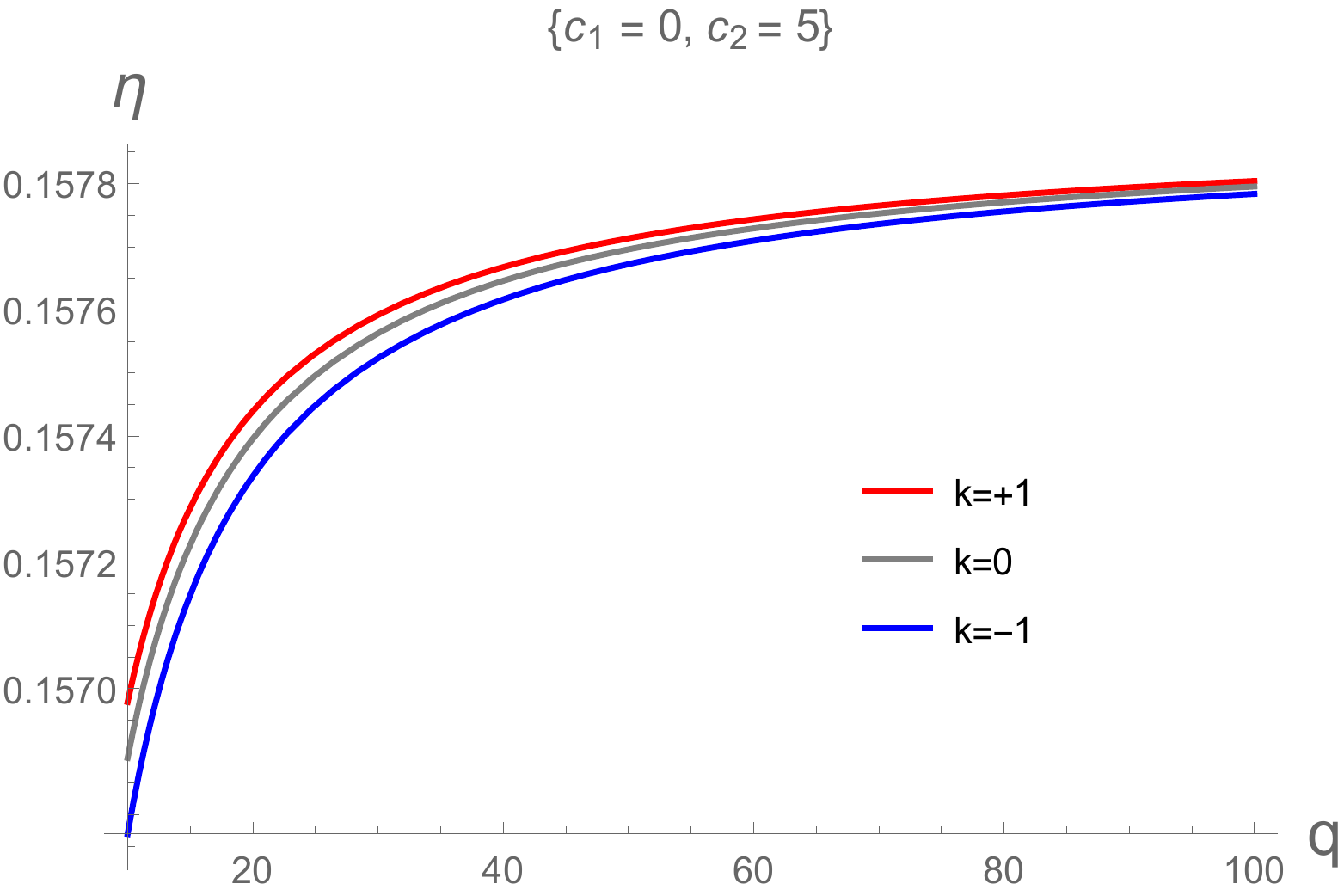}}\hspace{0.15cm}
 		\subfloat[]{\includegraphics[width=2in]{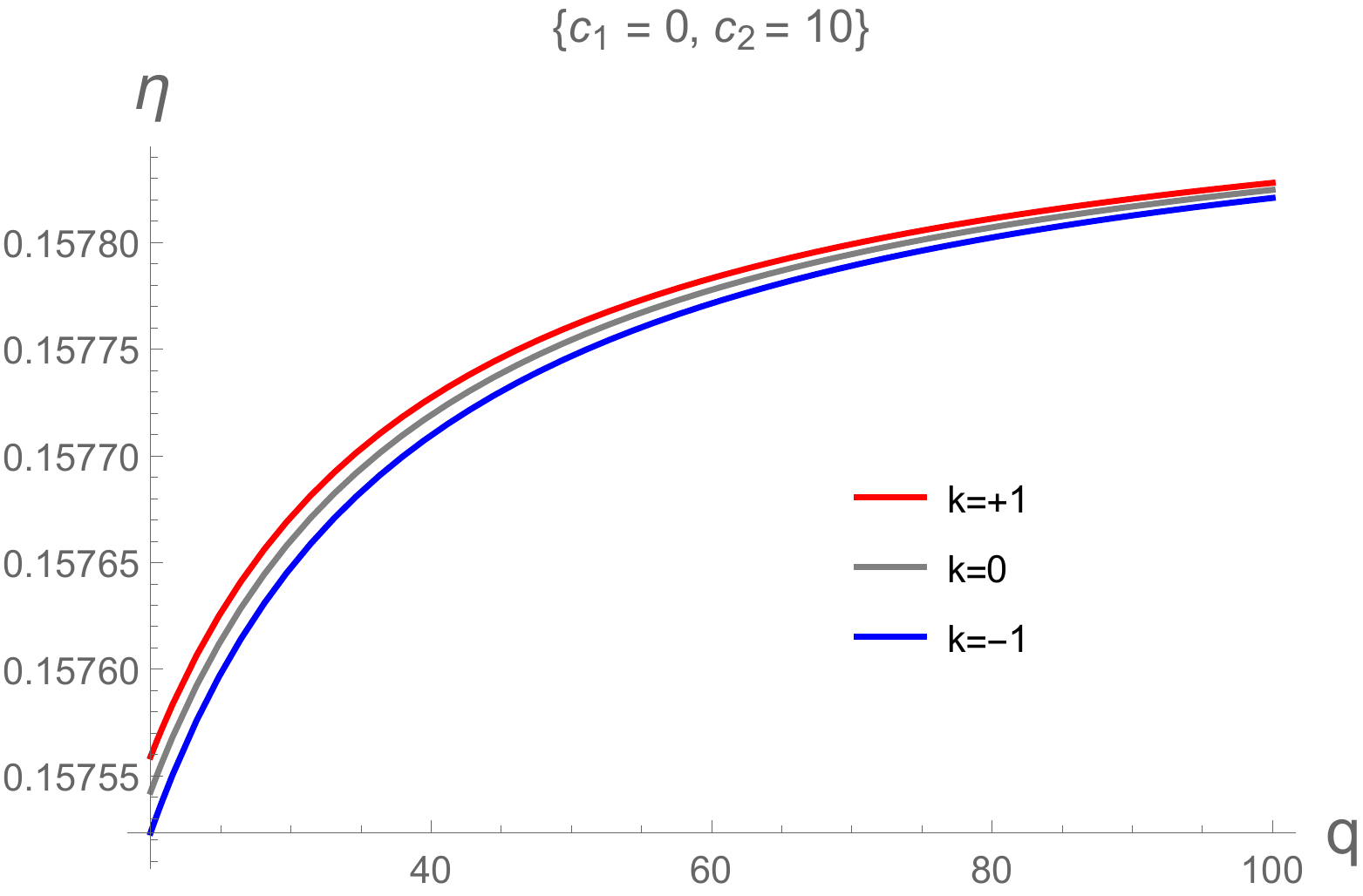}}

 		\caption{\footnotesize In the case of massive coefficient $c_1=0$, the behavior of efficiency $\eta$ with topology $k$ at various $c_2$ (a) $c_2 = 2$, (b) $c_2 = 5$ and (c) $c_2 = 10$. (Here, the parameters $L= m =c_0=1$,  are used.) 
 		} \label{fig:c2_sensitive} 
 	}
 \end{figure}

\begin{figure}[h!]
 	{\centering
 		\subfloat[]{\includegraphics[width=2in]{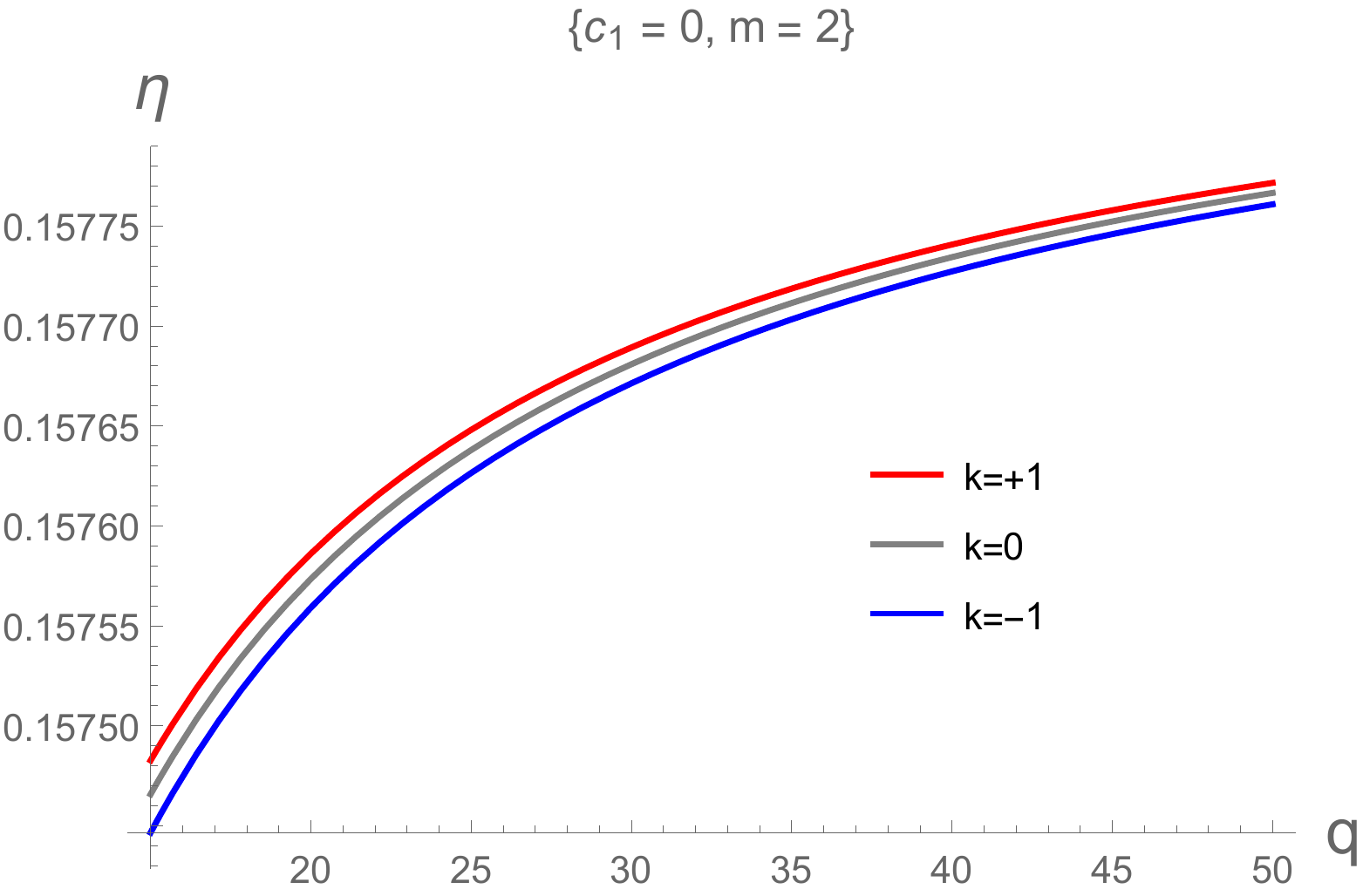}}\hspace{0.15cm}
 		\subfloat[]{\includegraphics[width=2in]{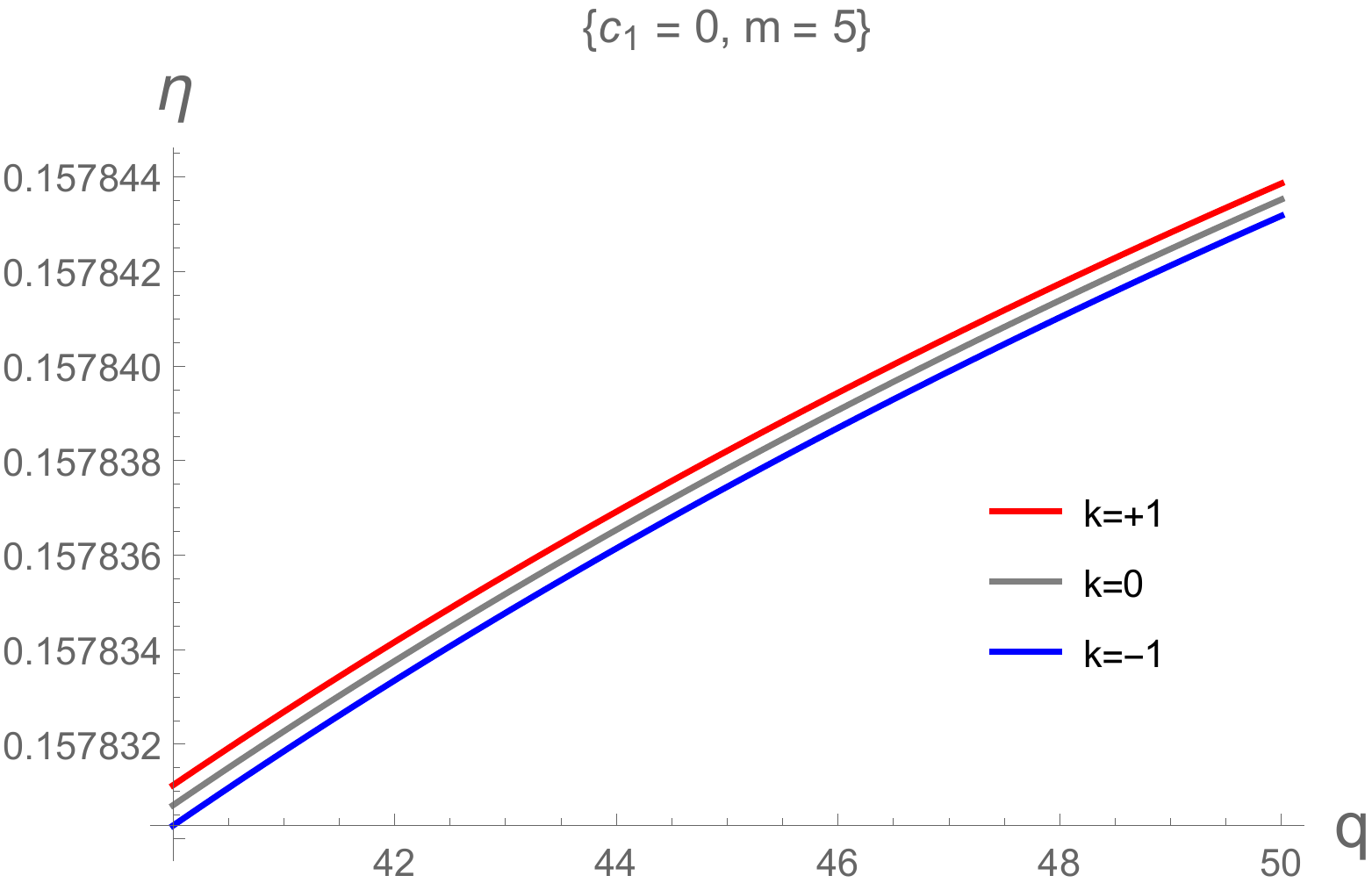}}\hspace{0.15cm}
 		\subfloat[]{\includegraphics[width=2in]{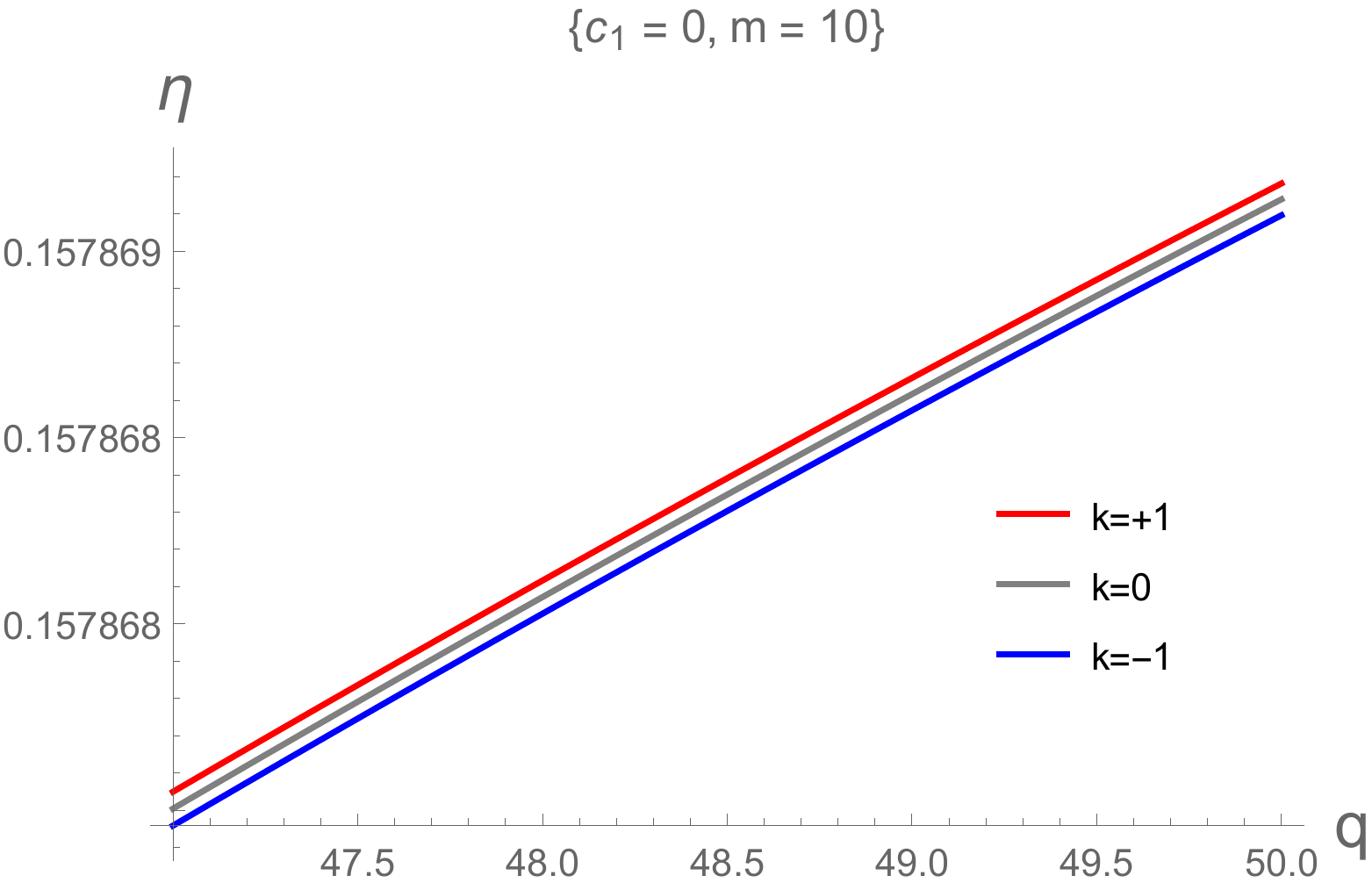}}

 		\caption{\footnotesize In the case of massive coefficient $c_1=0$, the behavior of efficiency $\eta$ with topology $k$ at various $m$ (a) $ m = 2$, (b) $ m = 5$ and (c) $ m = 10$. (Here, the parameters $L= c_0 =1$, and  $c_2 =3$, are used.) 
 		} \label{fig:m_sensitive} 
 	}
 \end{figure}

\noindent
First we concentrate on the case when the massive coefficient $c_1 = 0$. The heat inflow $Q_H$ can be obtained from eqn. \eqref{QH} in the binomial expansion (Since $0 < \frac{L}{q\sqrt{\epsilon}} < 1$) to be:
\begin{equation}\label{eq:QH approx}
Q_H \approx \frac{19L}{12\sqrt{6}} + \frac{2}{9\sqrt{6}}\frac{L^2}{q\sqrt{k+m^2 c_2 c_0^2}} \, .
\end{equation}
In this case, one can see that as the topological parameter $k$ increases from $k=-1$ to $+1$,
$Q_H$ decreases. The implication is that the efficiency $\eta$ increases (since, work is fixed) as $k$ takes higher values, i.e., one gets the order of efficiencies w.r.t. to topology as:
\begin{equation} \label{etaorder}
\eta_{\rm -1}^{\phantom{-1}} < \eta_{\rm 0}^{\phantom{0}} < \eta_{\rm +1}^{\phantom{+1}} .
\end{equation}
\noindent
Let us further note from the equations~\eqref{eq:work} and~\eqref{eq:QH approx}, that the result on the order of efficiencies in eqn.~\eqref{etaorder} goes over to the universal value $\eta \rightarrow \frac{3}{19}$, irrespective of topology $k$, in limit of $c_0$ (or $c_2$, or $q$, or $m$) $ \rightarrow \infty $. Where as, for any finite values of these parameters ($c_0, \ c_2, \ q, \ m$), the efficiency $\eta$ is as shown in figures~\eqref{fig:c0_sensitive},~\eqref{fig:c2_sensitive}, ~\eqref{fig:m_sensitive} and holds  the order found in eqn.~\eqref{etaorder}.
\vskip 0.4cm   \noindent
Hence, black holes with spherical topology are more efficient followed by flat topology and the black holes with hyperbolic topology are less efficient. This result is quite unexpected and in fact opposite to the result noted for heat engines in general topological black holes, when the thermodynamic cycle is considered far from criticality~\cite{Hennigar:2017apu}. In particular, the above topological order of efficiencies is also reversed compared to the case studied for massive gravity system in~\cite{Hendi:2017bys}. One of course needs to check how the ratio $\eta/\eta_C$ behaves too. For this, the behavior of Carnot  efficiency $\eta_{\rm C}^{\phantom{C}}$~\eqref{eq:itac} with topology can be estimated from:
\begin{equation}
\eta_{\rm C}^{\phantom{C}} \approx \frac{3}{19} + \frac{8}{513}\frac{L^3}{q^3(k+m^2 c_2 c_0^2)^{3/2}} \ ,
\end{equation}
which shows, $ \eta_{{\rm C}^{\phantom{C}}_{-1}} > \eta_{{\rm C}^{\phantom{C}}_{0}} > \eta_{{\rm C}^{\phantom{C}}_{+1}}$. Moreover, one can go for large charge limit as done in~\cite{Johnson:2017hxu}, to approach the Carnot limit.
 
\begin{figure}[h!]
	{\centering
		\subfloat[]{\includegraphics[width=3in]{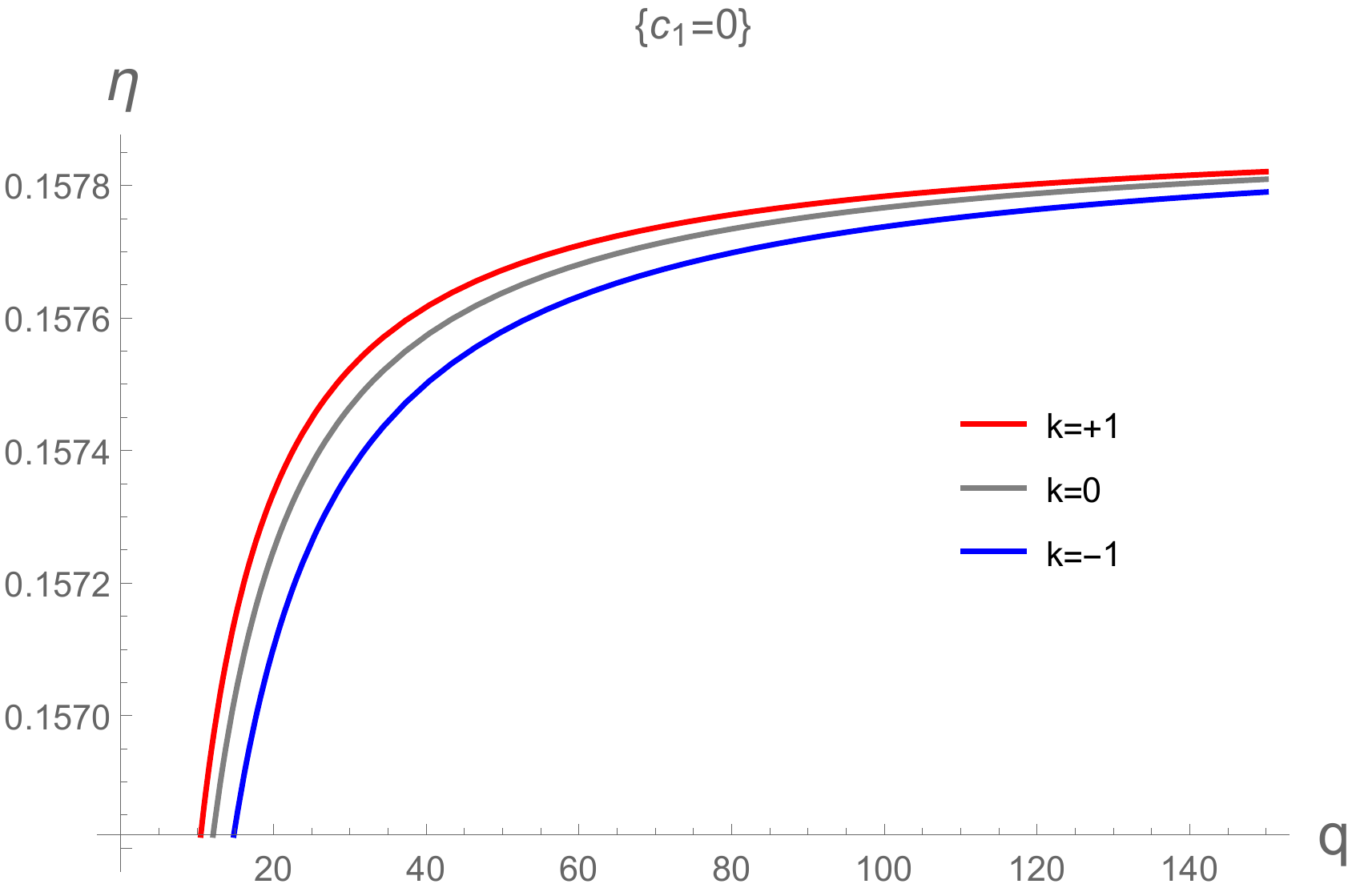}}\hspace{3cm}
		\subfloat[]{\includegraphics[width=3in]{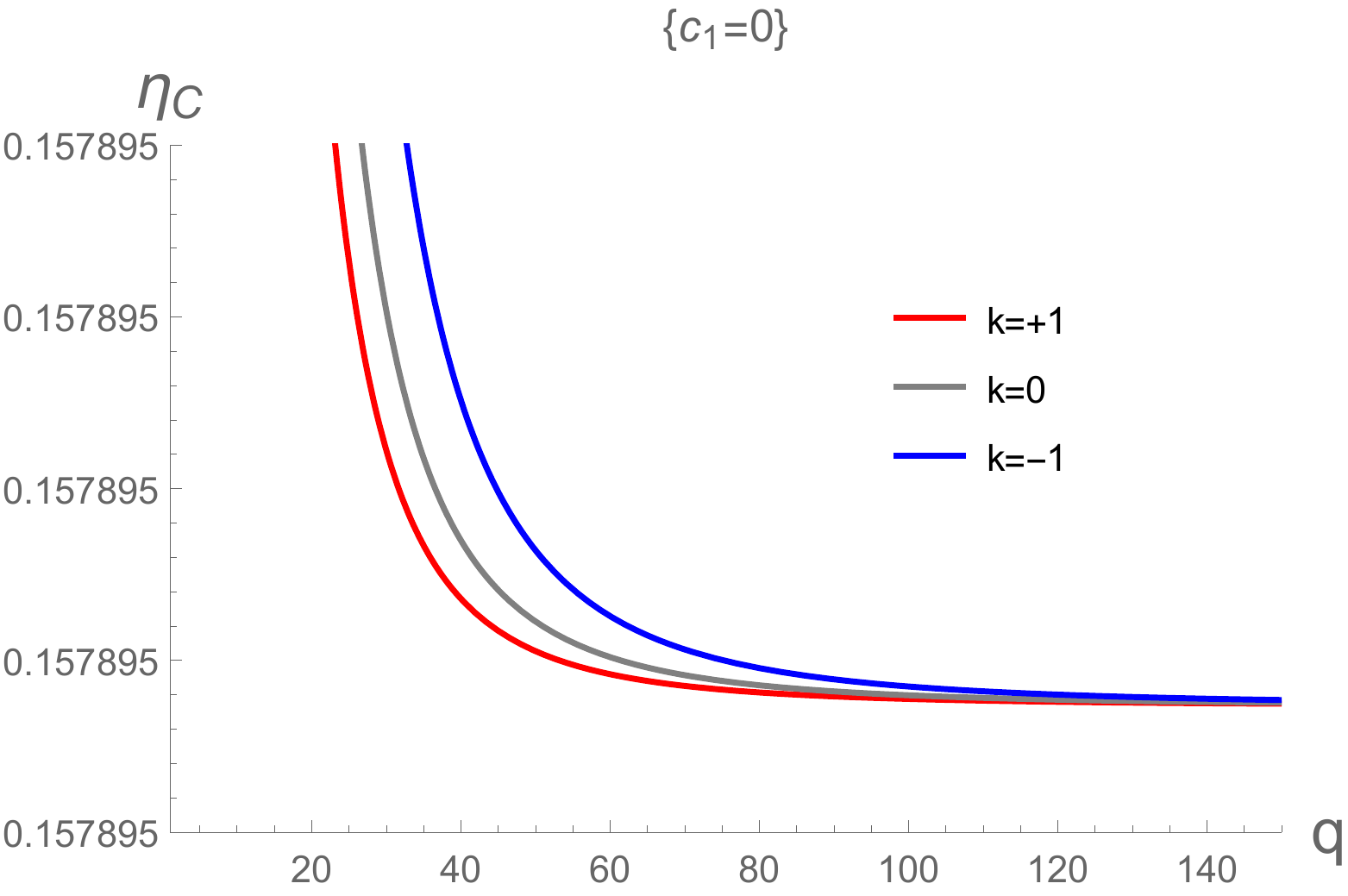}}\hspace{0.3cm}
		\subfloat[]{\includegraphics[width=3in]{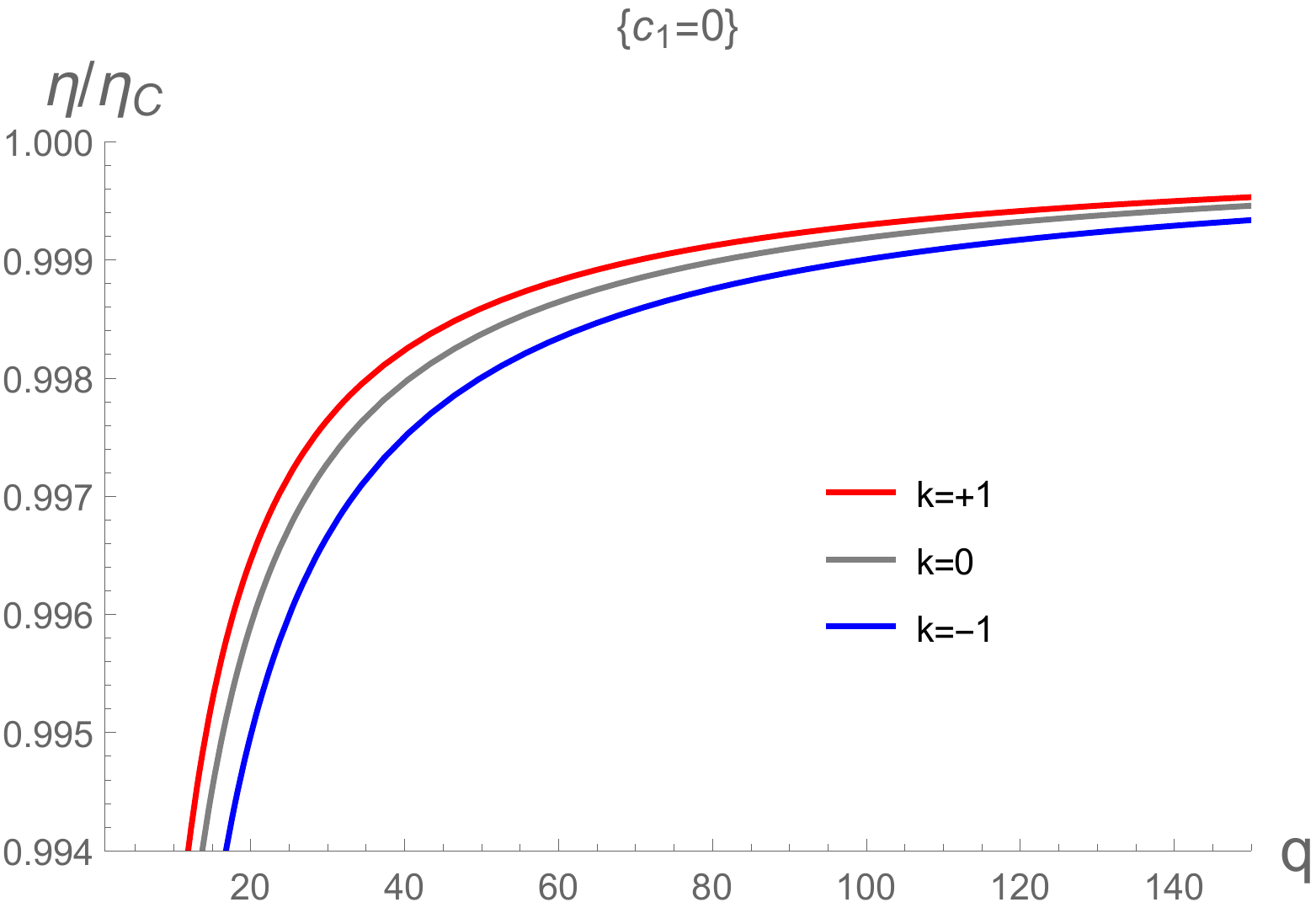}}				
		\caption{\footnotesize In the case of massive coefficient $c_1=0$, the effects of topology $k$ and charge $q$ on (a) $\eta$, (b) $\eta_{\rm C}$ and (c) $\eta/\eta_{\rm C}$. (Here, the parameters $L=m=c_0=1$, and  $c_2 =3$, are used.) 
			}   \label{fig:c1=0 plots}
	}
\end{figure}
\begin{figure}[h!]
	{\centering
		\subfloat[]{\includegraphics[width=2.6in]{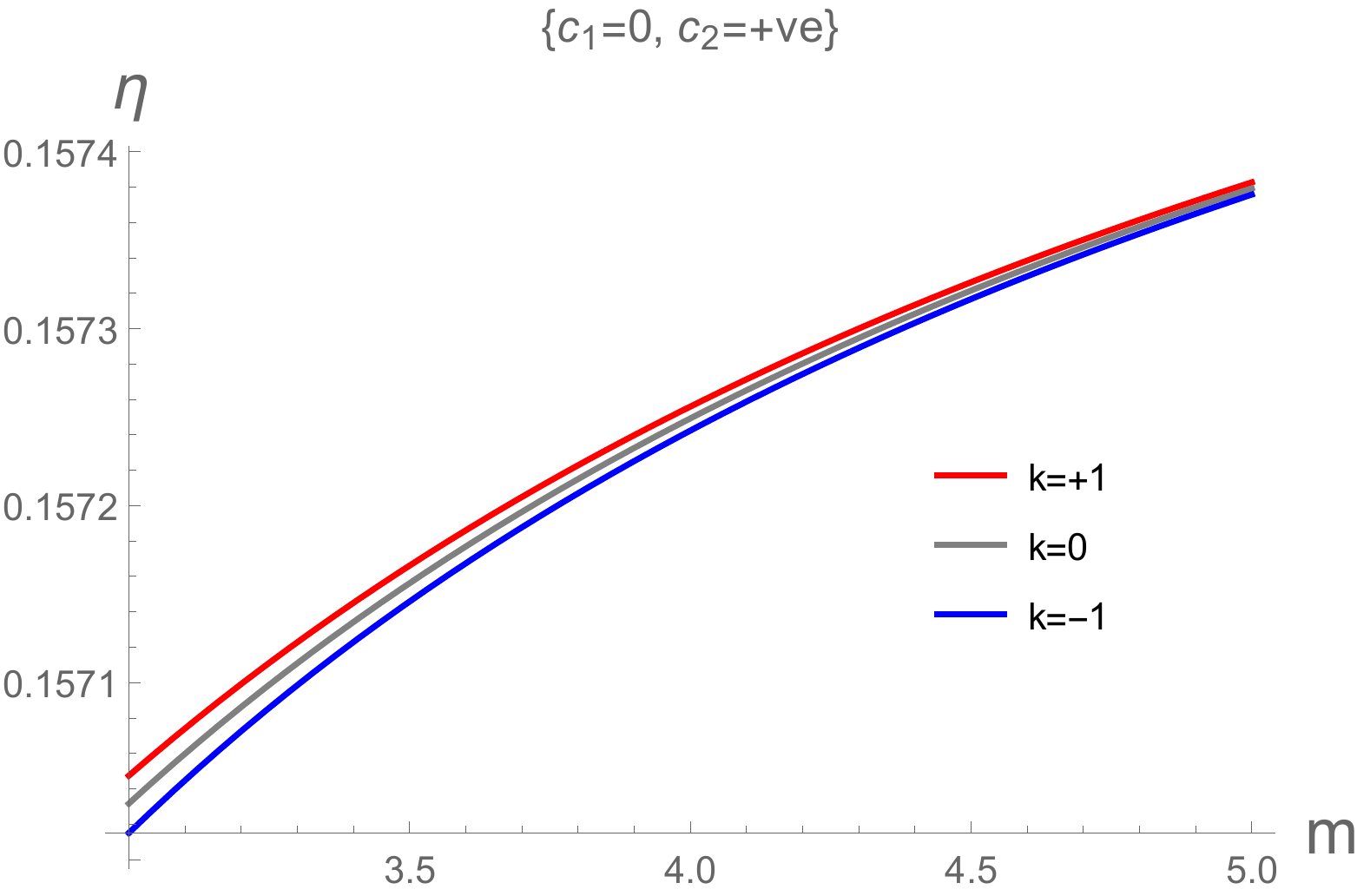}}\hspace{3cm}
		\subfloat[]{\includegraphics[width=2.6in]{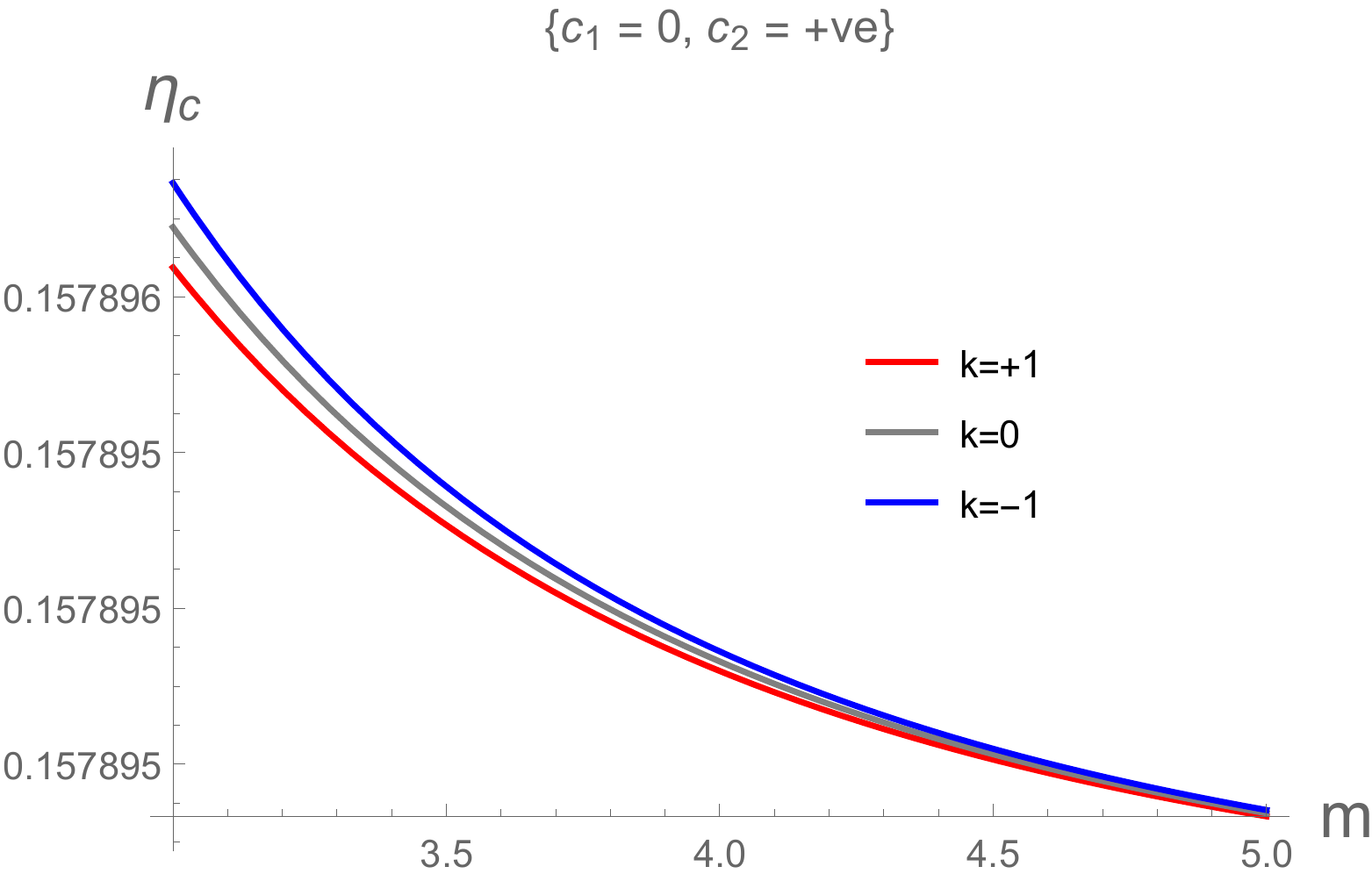}}\hspace{2cm}
		\subfloat[]{\includegraphics[width=2.6in]{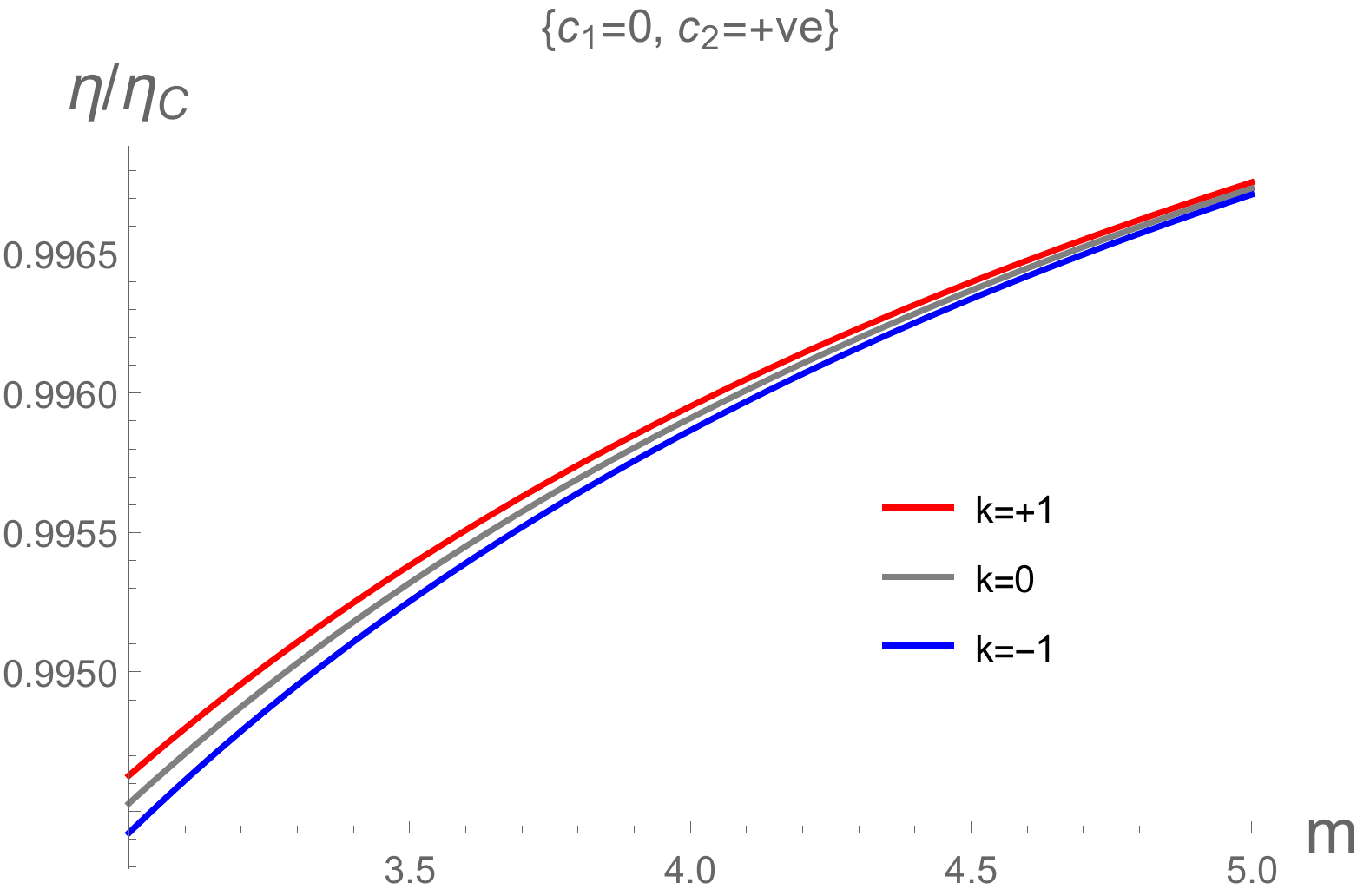}}

		\caption{\footnotesize  In the case of massive coefficients $c_1=0$, and positive ($+$ve) $c_2$, the effect of graviton mass $m$ on (a) $\eta$ (b) $\eta_C$ (c) $\eta/\eta_C$. (Here the parameters $c_0 =L=1, \, q=5, \, c_2=3$ are used).} \label{fig:meffect_k_plots}
	}
\end{figure}

\begin{figure}[h!]
	{\centering
		\subfloat[]{\includegraphics[width=2.35in]{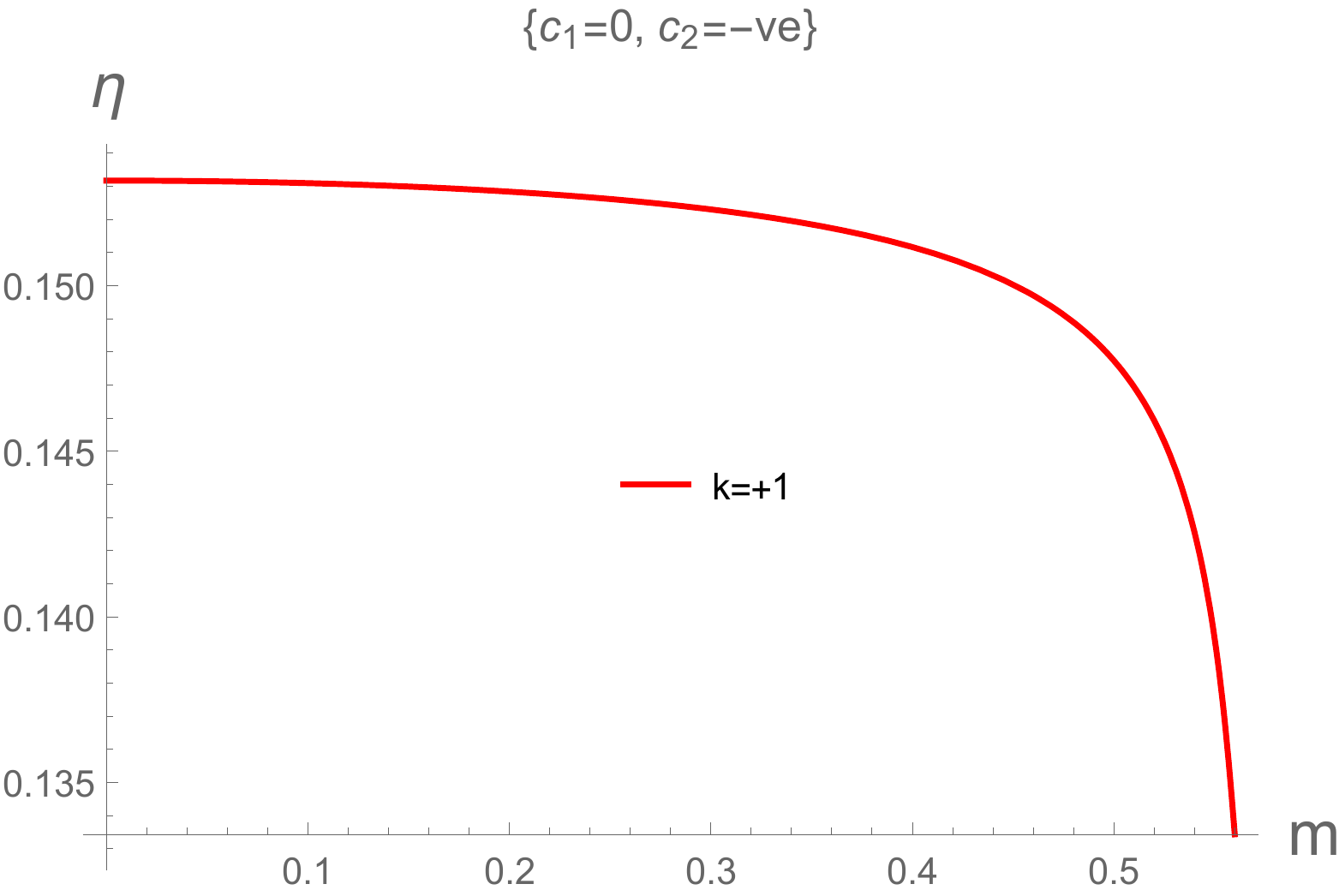}}\hspace{6cm}
		\subfloat[]{\includegraphics[width=2.35in]{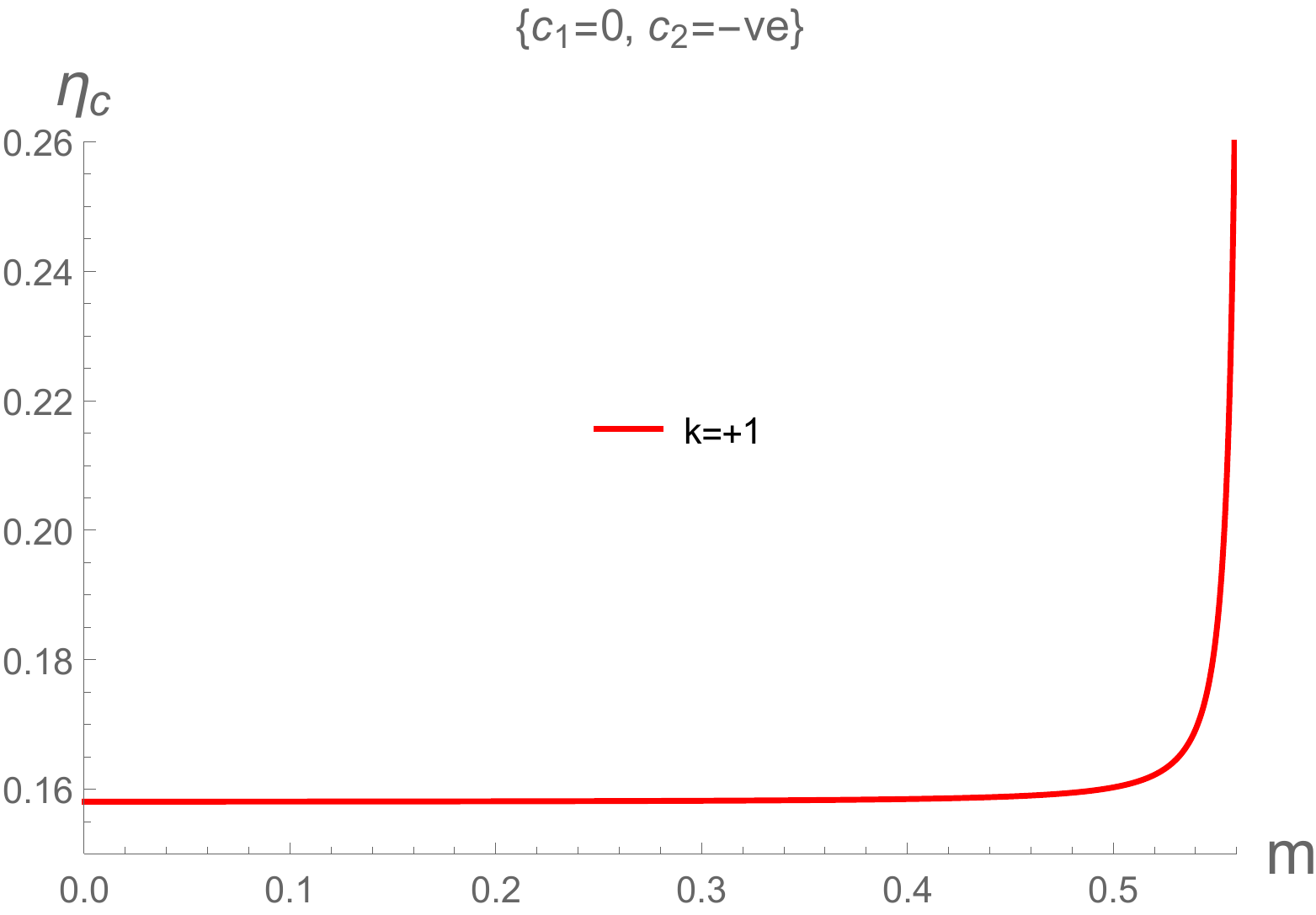}}\hspace{3cm}
		\subfloat[]{\includegraphics[width=2.35in]{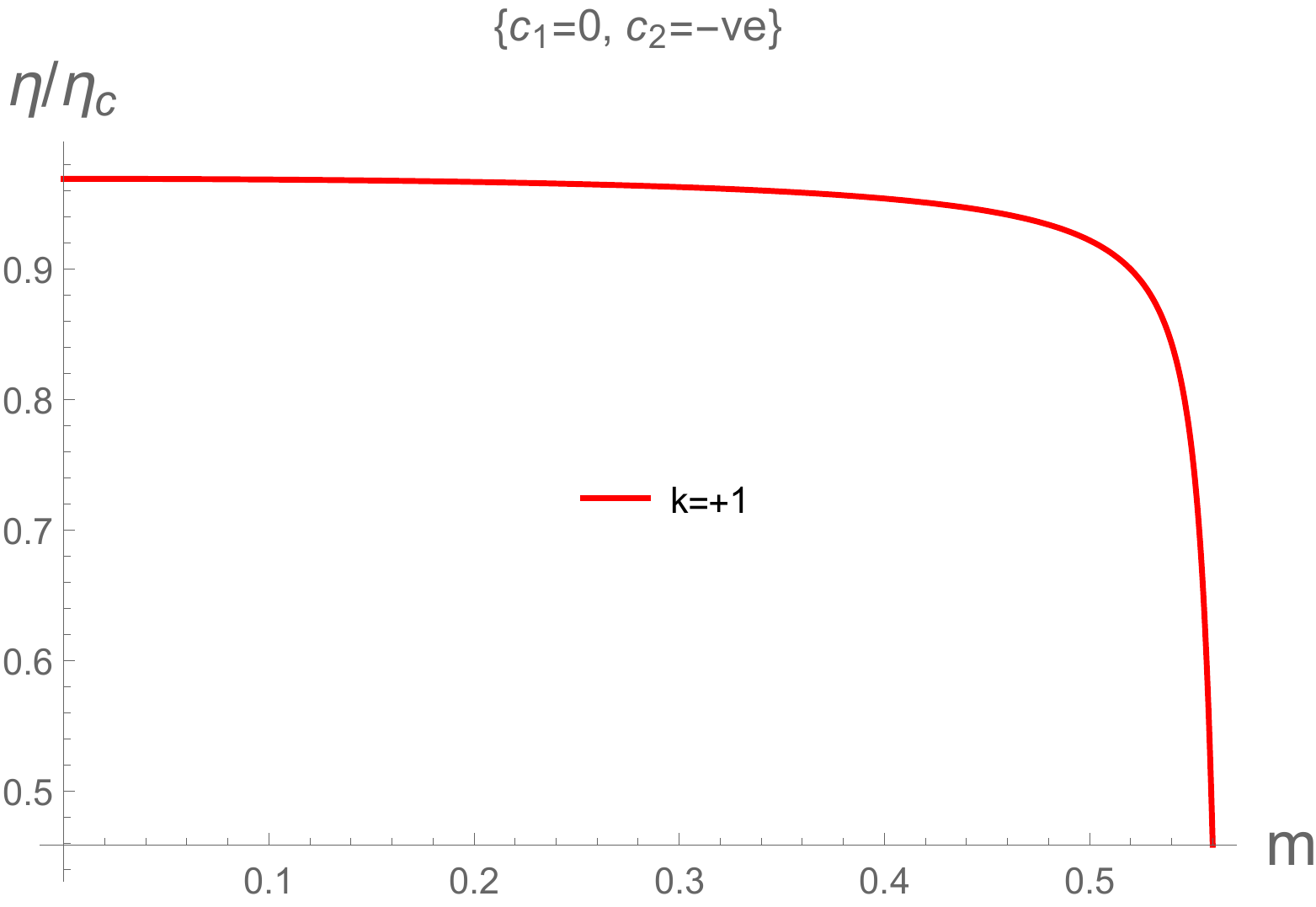}}

		\caption{\footnotesize  In the case of massive coefficients $c_1=0$, and negative ($-$ve) $c_2$, the effect of graviton mass $m$ on (a) $\eta$ (b) $\eta_C$ (c) $\eta/\eta_C$. (Here the parameters $c_0 =L=1, \, q=5, \, c_2=-3$ are used).  Note here that  when   $c_2$ is negative,  no critical behavior exist for $k= -1$ and $0$.} \label{fig:meffect_k_1_plots}
	}
\end{figure}
\begin{figure}[h!]
	{\centering
		\subfloat[]{\includegraphics[width=3.0in]{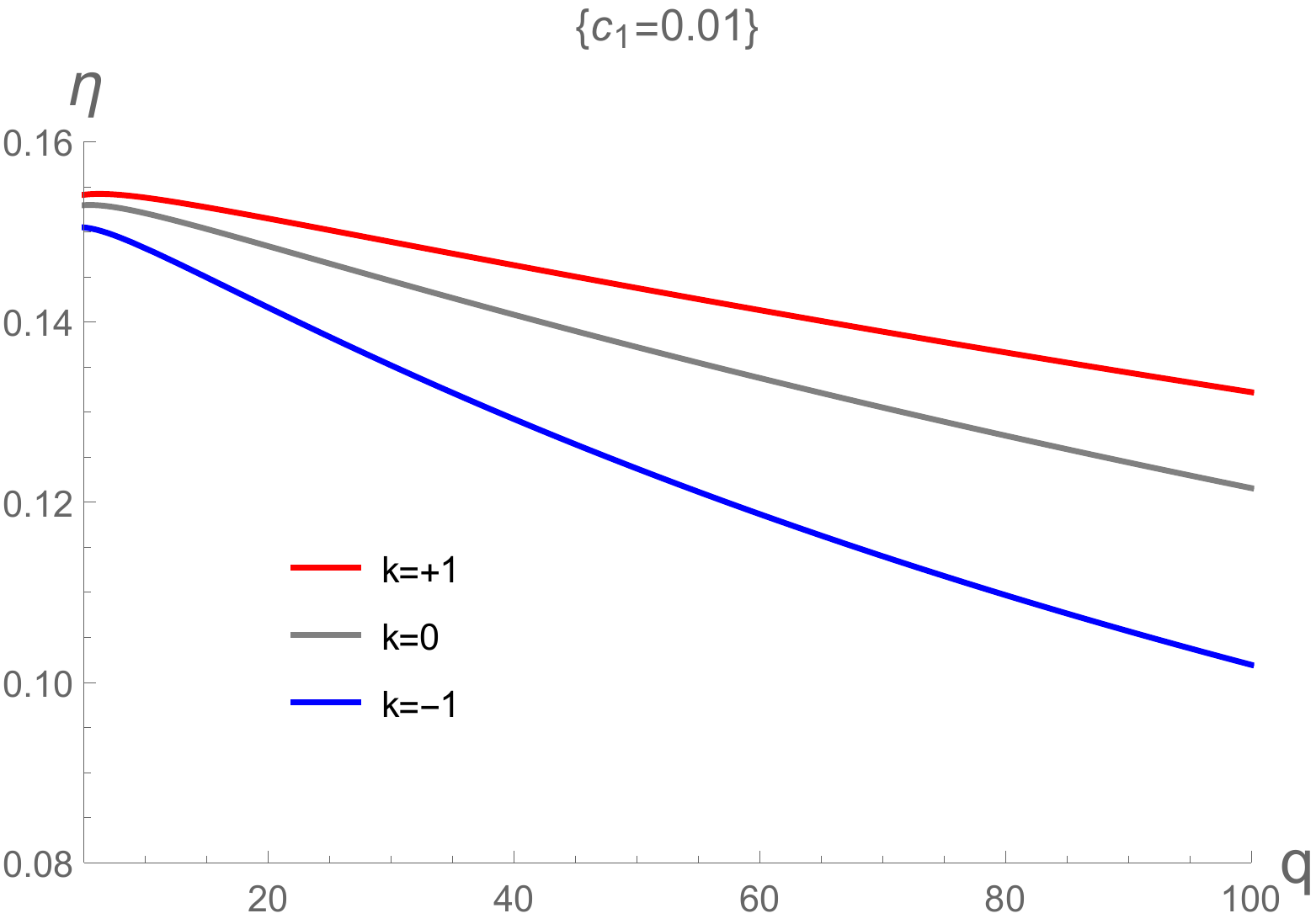}}\hspace{3cm}
		\subfloat[]{\includegraphics[width=3.0in]{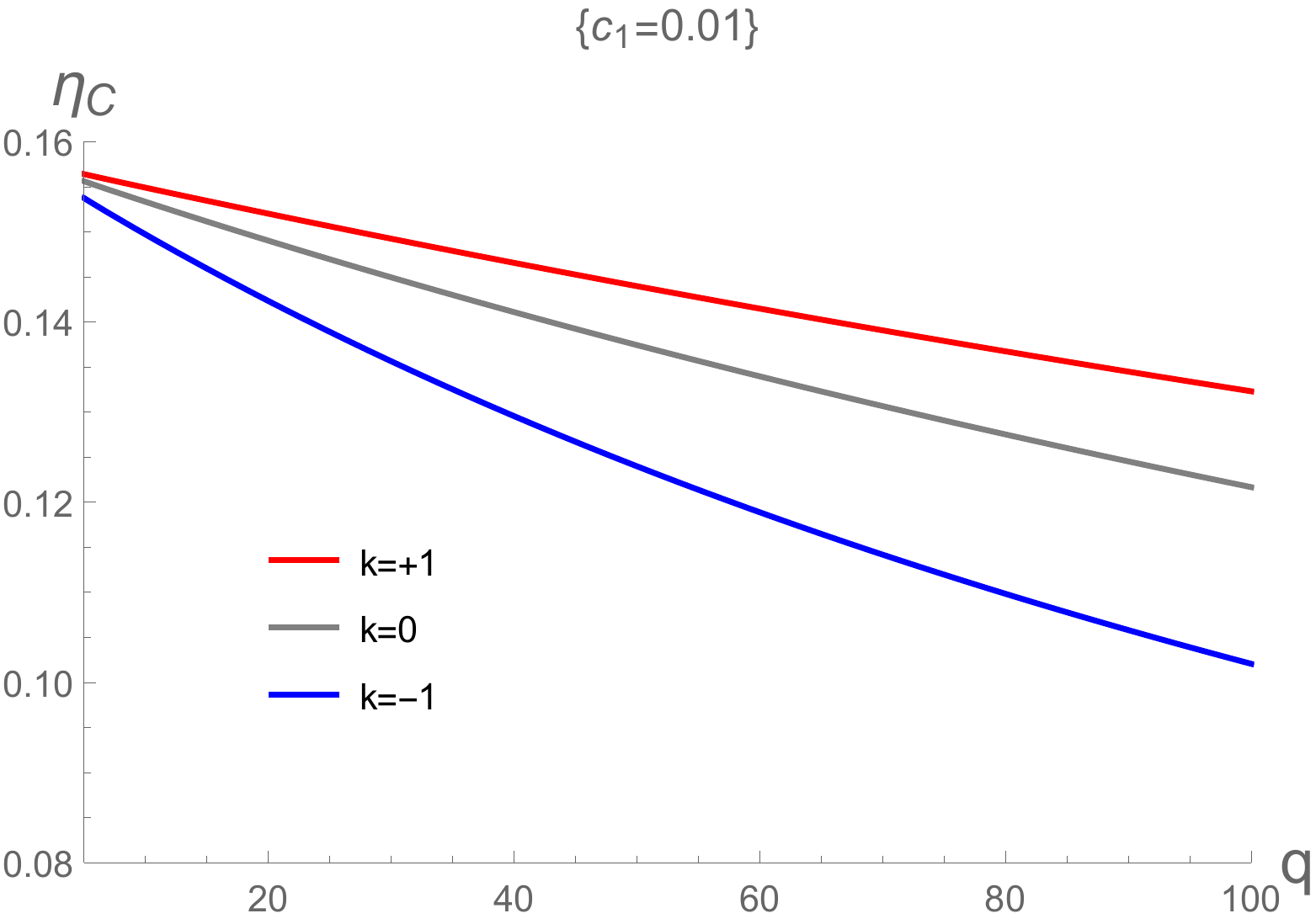}}\hspace{0.3cm}
		\subfloat[]{\includegraphics[width=3.0in]{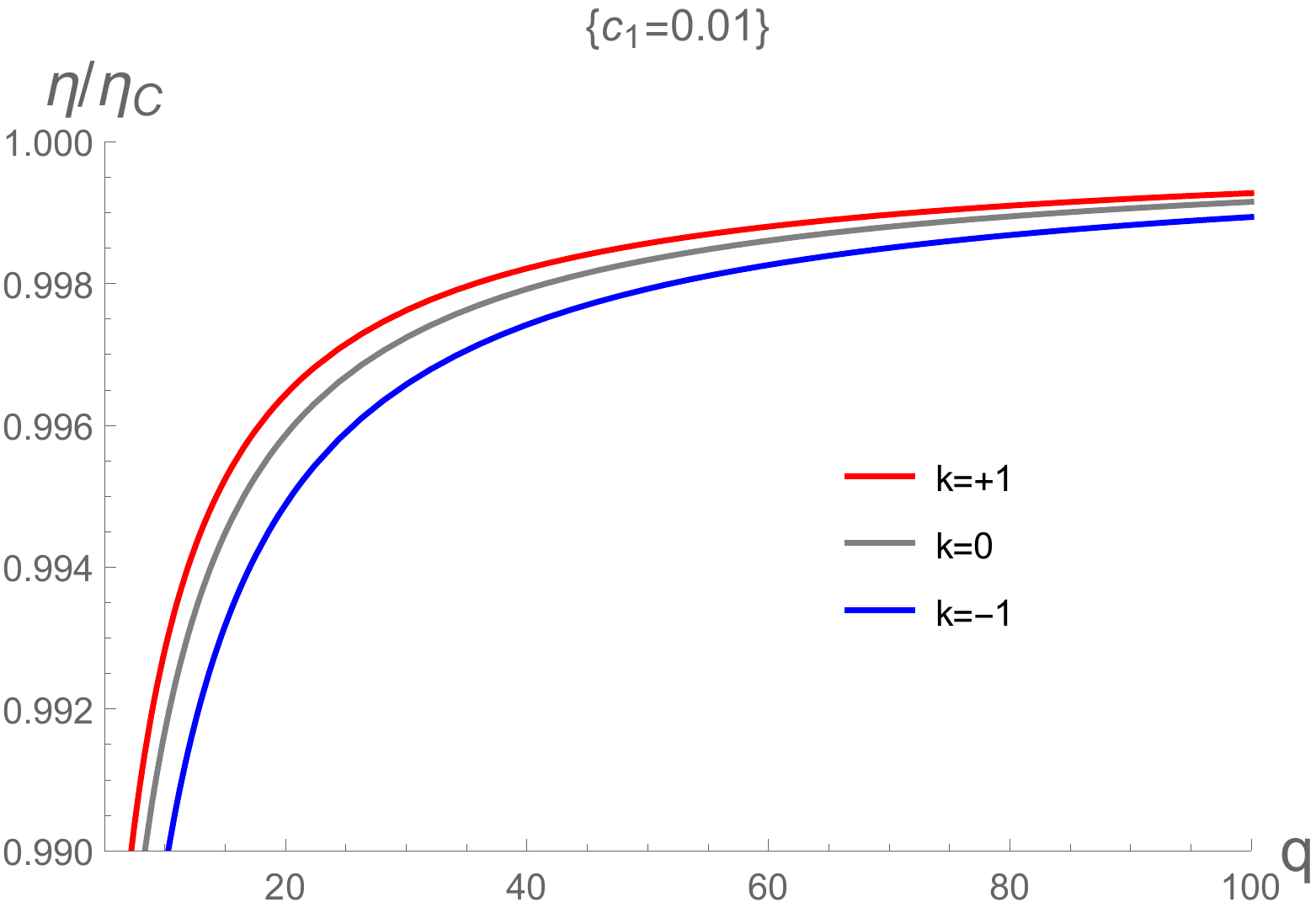}}\hspace{1cm}
		\caption{\footnotesize In the case of massive coefficient $c_1 \neq 0$, the effect of topology $k$  on  $\eta$,  $\eta_{\rm C}$ and  $\eta/\eta_{\rm C}$, over a sample range of charge $q$ and $c_1=0.01$ in (a), (b), (c). (Here, the parameters $L=m=c_0=1$, and  $c_2 =3$, are used.) 
		}   \label{fig:c1p non zero plots}
	}
\end{figure}
\begin{figure}[h!]
	{\centering
		\subfloat[]{\includegraphics[width=3.0in]{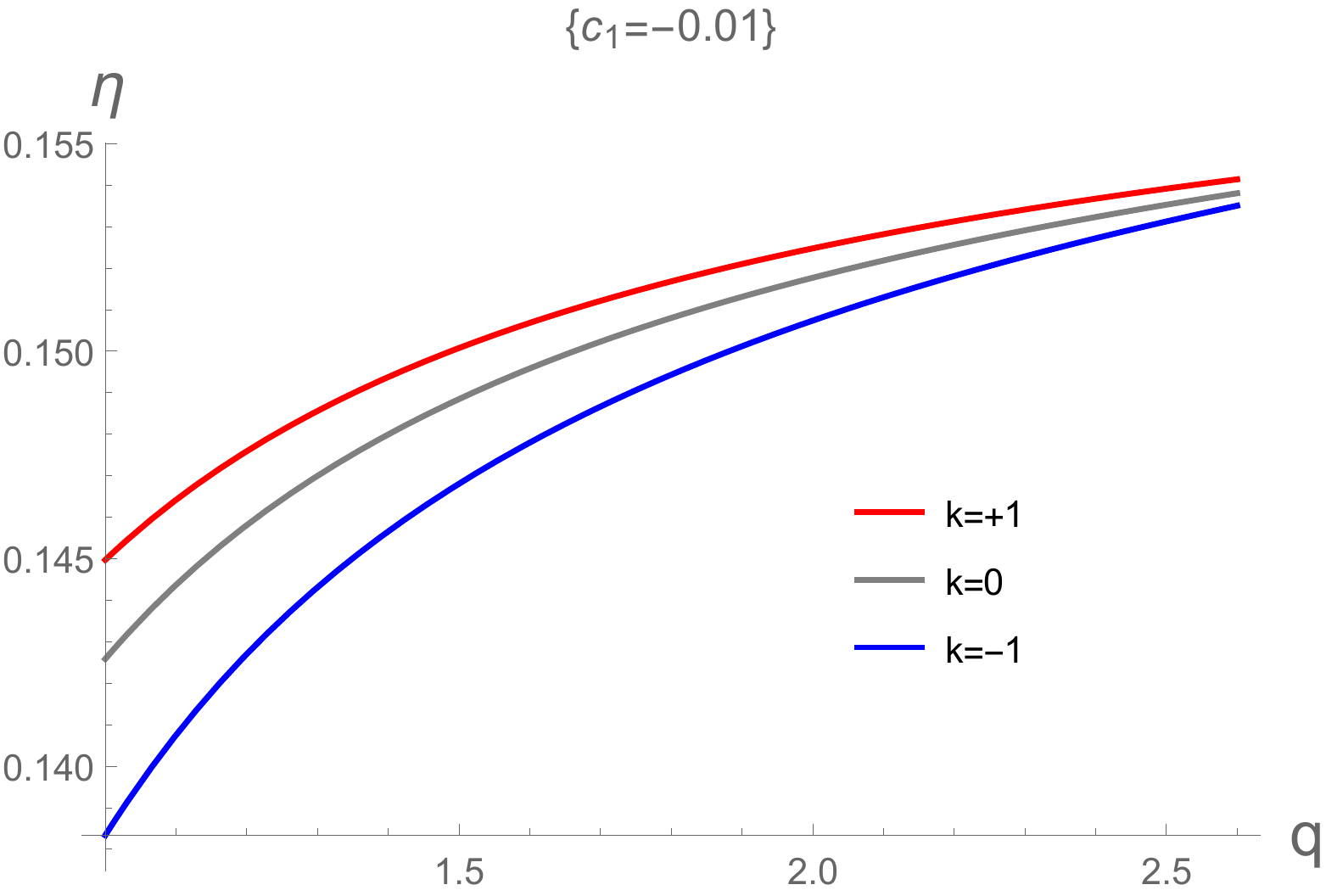}}\hspace{3cm}
		\subfloat[]{\includegraphics[width=3.0in]{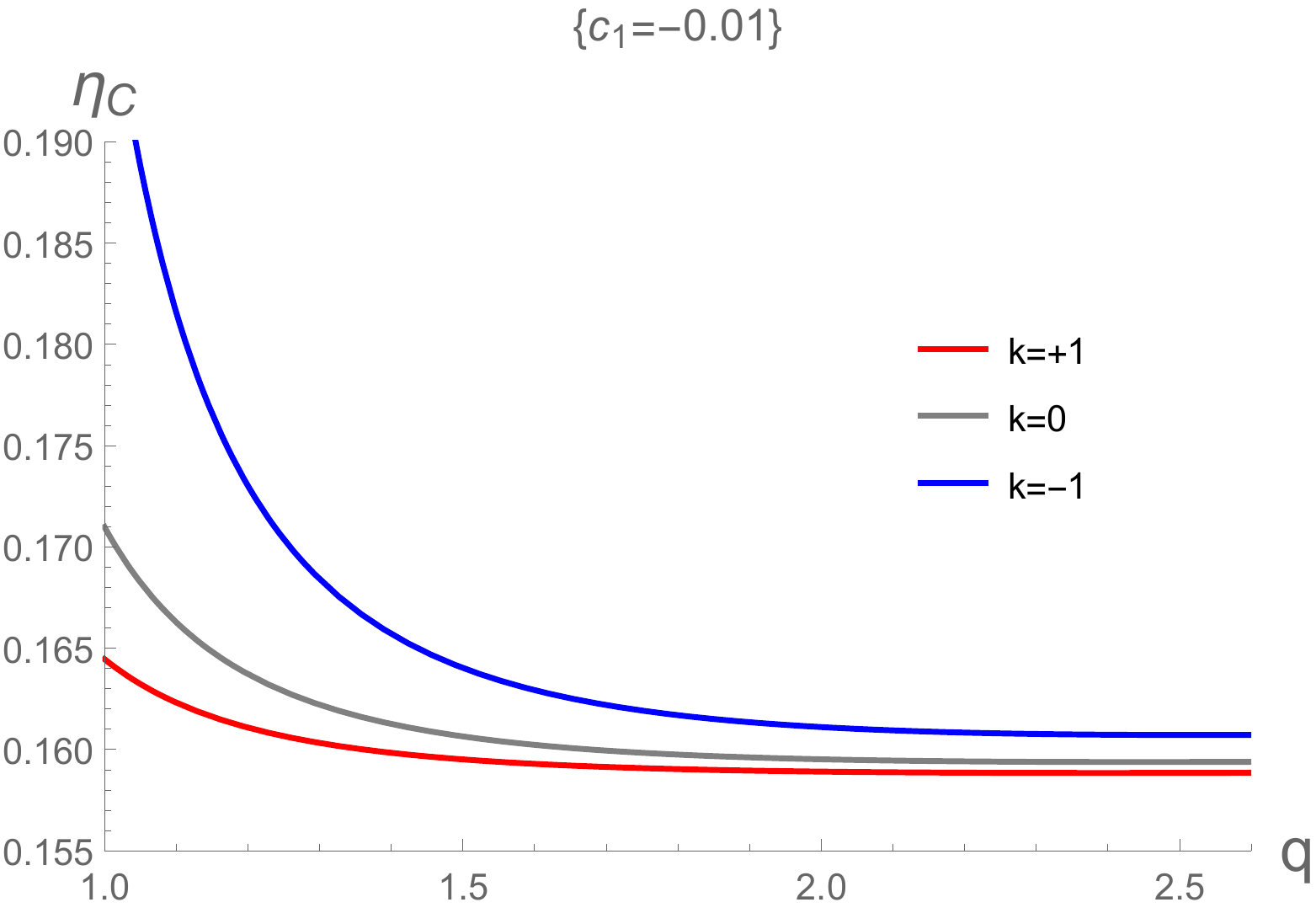}}\hspace{0.3cm}
		\subfloat[]{\includegraphics[width=3.0in]{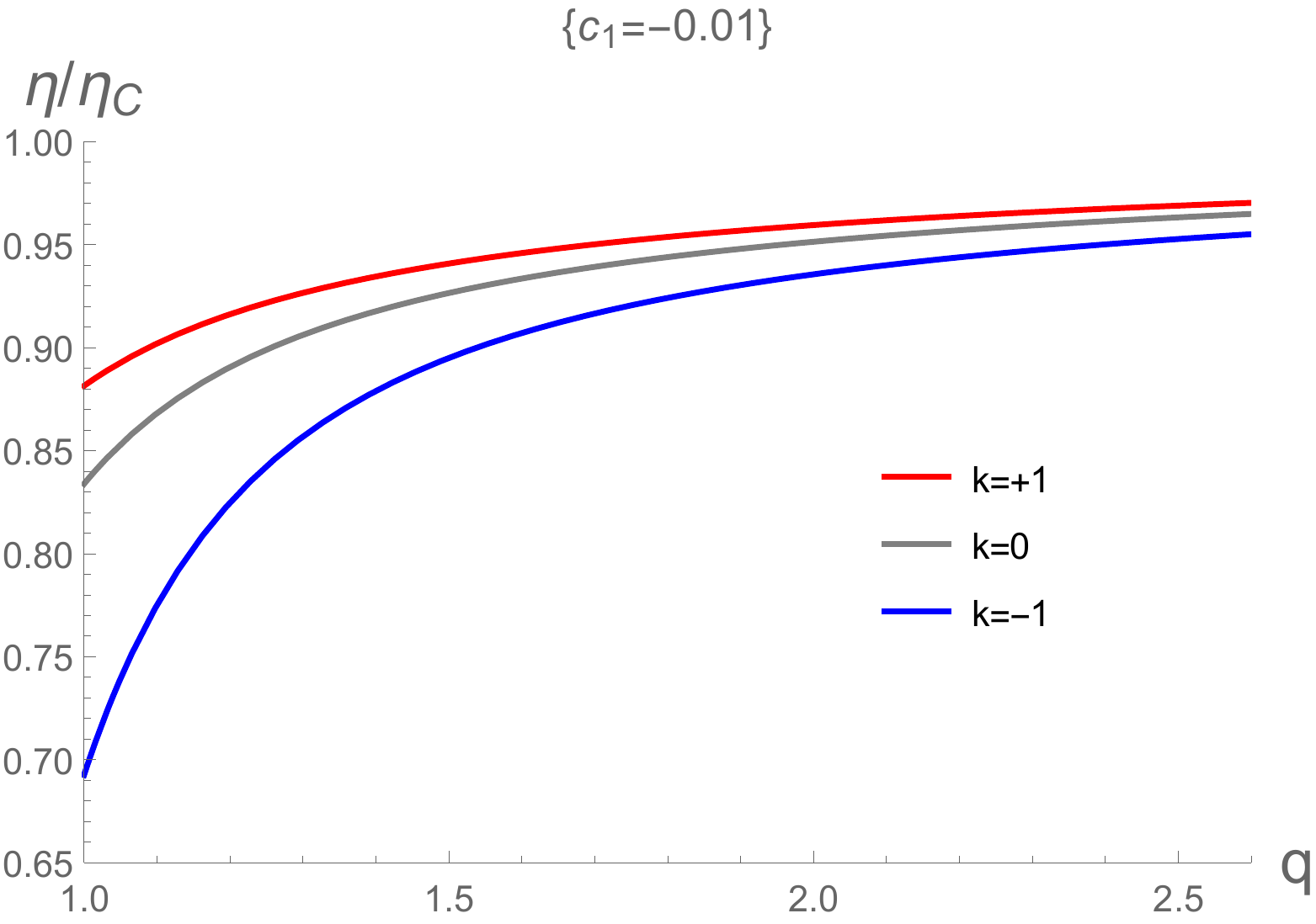}}	
		\caption{\footnotesize In the case of massive coefficient $c_1 \neq 0$, the effect of topology $k$  on  $\eta$,  $\eta_{\rm C}$ and  $\eta/\eta_{\rm C}$, over a sample range of charge $q$ and $c_1= - 0.01$ in (a), (b), (c). (Here, the parameters $L=m=c_0=1$, and  $c_2 =3$, are used.) 
		}   \label{fig:c1n non zero plots}
	}
\end{figure}
In the large charge limit, the expressions for $Q_H, \eta$ and $\eta_{\rm C}^{\phantom{C}}$ are given by:
\begin{eqnarray} \label{QHq}
Q_H  &=& \frac{19\sqrt{6}}{72}L+\frac{\sqrt{6}}{27}{\frac {{L}^{2}}{q\sqrt{\epsilon}}}+\frac{4\sqrt{6}}{243}{\frac {{L}^{3}}{q^2\epsilon}}+\frac {25\sqrt{6}}{2916}\frac{L^4}{q^3\epsilon^{3/2}} + O \left( q^{-4} \right)\ , \\ \label{etaq}
\eta &=& \frac{3}{19} -\frac{8}{361}\frac{L}{q\sqrt{\epsilon}}-\frac{416}{61731}\frac{L^2}{q^2\epsilon}-\frac{3286}{1172889}\frac {L^3}{q^3\epsilon^{3/2}}-\frac {764594}{601692057}\frac{L^4}{q^4\epsilon^2} + O \big( q^{-5} \big) \ ,  \ \ \ \ \\ \label{etaCq}
\eta_{\rm C}^{\phantom{C}} &=& \frac{3}{19} +\frac{8}{513}\frac{L^3}{q^3\epsilon^{3/2}}+\frac{14}{513}\frac{L^4}{q^4\epsilon^2}+\frac{166}{4617}\frac {L^5}{q^5\epsilon^{5/2}}+\frac {590}{13851}\frac{L^6}{q^6\epsilon^3} + O \left( q^{-7} \right) \ .
\end{eqnarray}
As a consistency check of our results, we note that taking the special values $m=0$ and $k=1$ (corresponding to charged black holes in AdS with corrections from massive gravity dropped) the above expressions in eqn. (\ref{QHq})-(\ref{etaCq}) for large $q$, yield exactly the large $q$ results of Johnson~\cite{Johnson:2017hxu}. \\

\noindent
We can now see the  effect of topology and graviton mass on efficiency of our engines.
Figure~\eqref{fig:c1=0 plots} shows that, at fixed topology as the charge $q$ increases, the Carnot efficiency $\eta_{\rm C}$ decreases, while the efficiency $\eta$ and the ratio  $\eta/\eta_{\rm C}$ increase. However, $\eta= \eta_{\rm C}$ is possible only in the limit $q \rightarrow \infty$.
At finite charge, the behavior of the quantities, $\eta, \eta_{\rm C}$ and   $\eta/\eta_{\rm C}$, with topology is quite different, with the exception that in the limit  $q \rightarrow \infty$, their topological dependence vanishes.  Furthermore, as shown in figures~\eqref{fig:meffect_k_plots} and~\eqref{fig:meffect_k_1_plots},  the presence of graviton mass $m$ improves the efficiency $\eta$  when $c_2$ is positive,  while lowers the efficiency $\eta$ when $c_2$ is negative. \\

\noindent
For the case $c_1 \neq 0$,  figures~\eqref{fig:c1p non zero plots} and ~\eqref{fig:c1n non zero plots}, show that the variation of efficiency for various topologies, follows the inequality presented in eqn. (\ref{etaorder}), when the thermodynamic cycle is placed closed to critical point.

\subsection{Critical region of black holes}
\begin{figure}[h!]
	{\centering
		\includegraphics[width=4.0in]{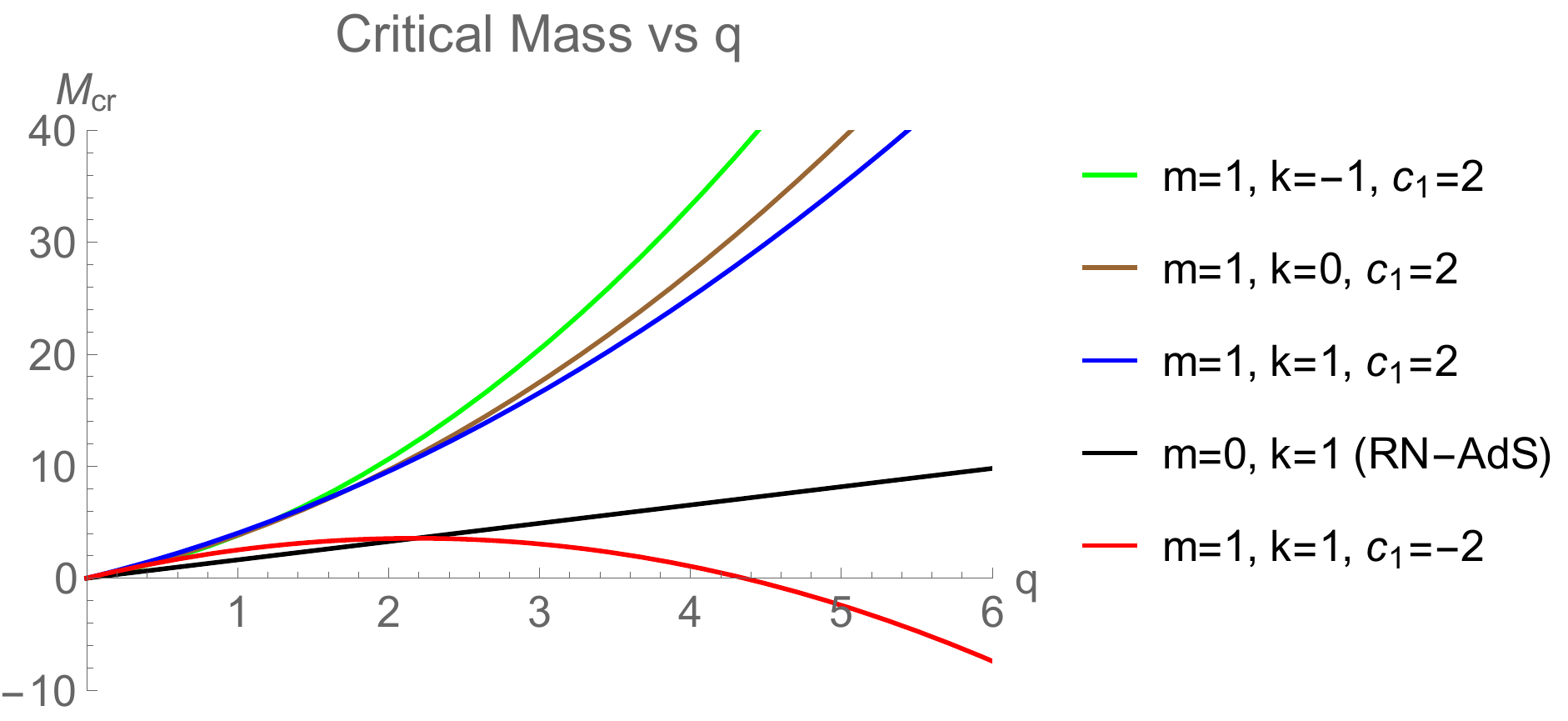}		
		\caption{\footnotesize   
			Variation of critical Mass w.r.t. charge q for various cases as compared to RN-AdS case.}   \label{fig:mcrit}
	}
\end{figure}

\begin{figure}[h!]
	{\centering
		\includegraphics[width=5.0in]{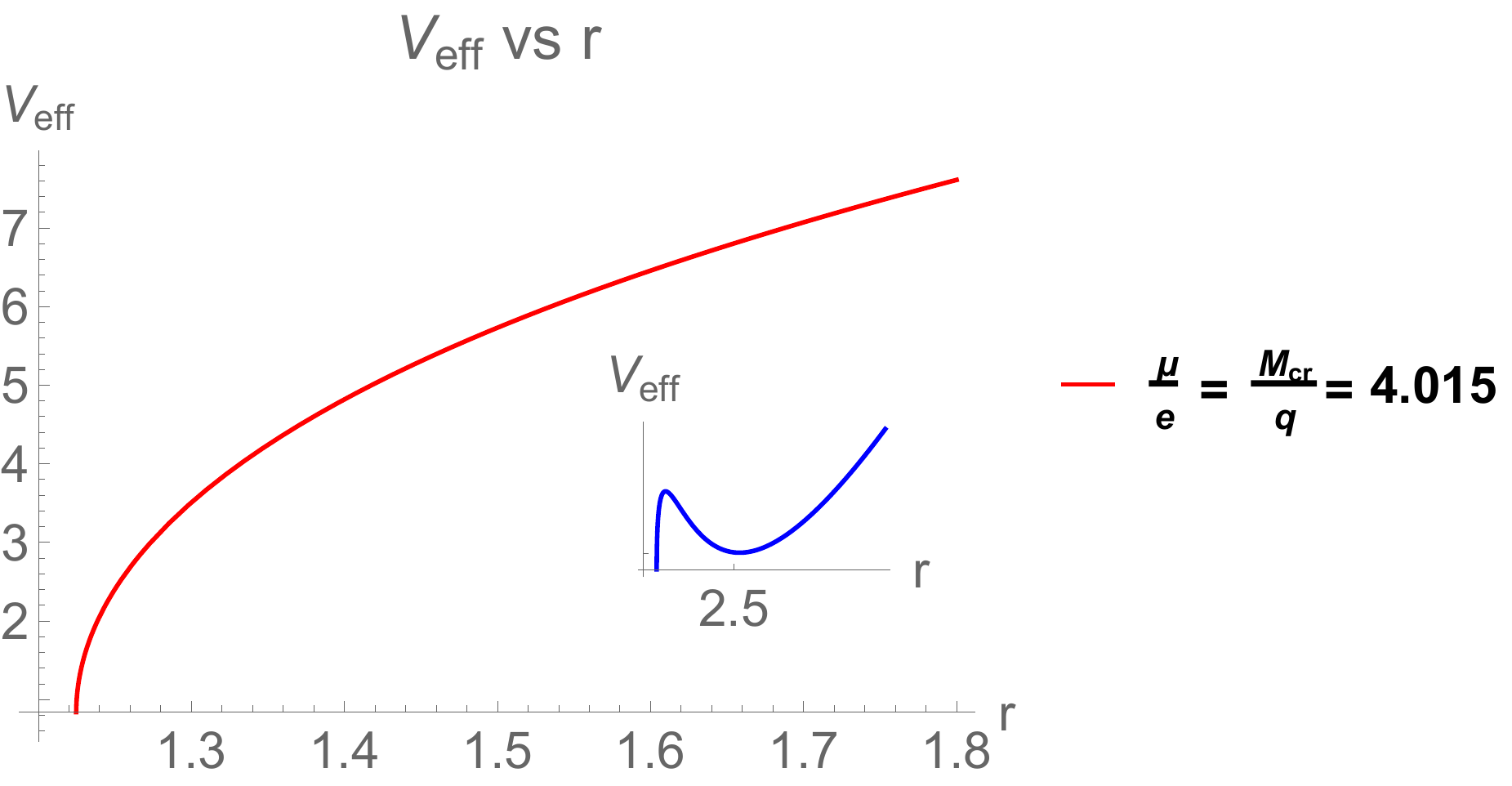}		
		\caption{\footnotesize   
			Main plot: Effective Potential for $L=0$, $k=q=1,m=1,c_0=1,c_2=2,c_3=3$. Inset: $\mu/e=0.1745, M_{cr}/q=4.015$.}   \label{fig:eff}
	}
\end{figure}

We now move on to study the critical region of the black hole, following the idea that a large number of coupled microscopic subsystems can drive the system to Carnot efficiency at criticality~\cite{PhysRevLett.114.050601,power_of_a_critical_heat}. In the context of black holes, this can be done by using a toy model of $q$ interacting constituent objects~\cite{Johnson:2017hxu} in the background of critical hole. In particular, we consider a particle of mass $\mu$ moving in the background of this critical black hole in the probe approximation. Following the methods in~\cite{Johnson:2017asf,Chandrasekhar1984,doi:10.1142/S0217732311037261}, the effective potential is seen to be
\begin{equation} \label{eq:staticform}
V_{\rm eff}(r) = \frac{e\, q}{r}+\sqrt{Y_{\rm cr}(r)}\sqrt{\mu^2+\frac{L^2}{r^2}}\ ,
\end{equation}
with $L$ denoting the angular momentum of the particle. To take closer look at the critical region one studies the metric function of charged massive black hole geometry  (\ref{eq:metric}), with critical values inserted, i.e.,
\begin{equation}
Y_{\rm cr}(r) =k-\frac{2 M_{\rm cr}}{r}+\frac{ r^{2}}{l^2_{cr}}+\frac{q^{2}}{ r^{2}}+m^{2}(\frac{c_0c_{1}}{2}r+c_0^{2}c_{2}) \ ,
\label{eq:Ycrit}
\end{equation}
where
\begin{eqnarray} \label{Mcr}
&& M_{\rm cr} = \frac{q \left(3 \sqrt{6}c_0c_1 m^2 q \sqrt{\frac{1}{\left(c_0^2 c_2 m^2+k\right)^3}}+8\right)}{2 \sqrt{6}
   \sqrt{\frac{1}{c_0^2 c_2 m^2+k}}} \\ \nonumber
&& l^2_{\rm cr} =  36\,\frac{q^2}{\epsilon^2} \nonumber
\end{eqnarray}
The critical values of mass $M_{cr}$ and cosmological constant parameter $l_{cr}$, are closely related to the RN-AdS case studied in~\cite{Johnson:2017asf} with corrections involving massive gravity parameters. The critical mass is plotted in figure-(\ref{fig:mcrit}) for various values of two key parameters, namely, $k$ and $c_1$, in comparison to the RN-AdS case (where $M_{cr}$ is a linear function of charge $q$). In the present case, in the large charge limit, for any value of other parameters of the model (such as $k,m,c_0,c_2$):  $M_{cr}$ is always higher than the RN-AdS case for positive $c_1$ and can be less than RN-AdS case, including vanishing at some $q$, when $c_1$ is negative. This can be seen from the expression for critical mass in eqn. (\ref{Mcr}), plotted in figure-(\ref{fig:mcrit}), showing that $M_{cr}$ vanishes at two points, namely, at $q=0$ and at
\begin{equation} \label{qvanish}
q= -\frac{4 \sqrt{\frac{2}{3}} \left(c_0^2c_2 m^2+k\right)^{3/2}}{3c_0 c_1 m^2} \, .
\end{equation}
In contrast, in the RN-AdS case, $M_{cr}$ vanishes only when $q=0$. Implications of this feature while considering near horizon limit of critical black holes will be discussed below. 
First, we study the effective potential in eqn. (\ref{eq:staticform}), which may generally have a minimum at some value of $r_{min}$ ($> r_{cr}$), depending on the values taken by $\mu, e$ and $L$. It was argued in~\cite{Johnson:2017asf}, that the presence of such a local minimum for the critical hole would lead to a condensation and possibly an instability. The presence or absence of such a minimum can be studied numerically by looking for a possible $r_{min}$ with the mass to charge ratio taken to be identical to $M_{cr}/q$. As can be seen from figure (\ref{fig:eff}), there is a local minimum for certain values of $\mu/e$, which quickly disappears once $\mu/e=M_{cr}/q$ and hence the potential is purely attractive type binding all the microsystems together, with no local minimum.
Now, we can analyze another aspect of the critical black hole by taking a double scaling limit
where the charge parameter $q$ is taken to be large while at the same time nearing the horizon. The analysis can proceed parallel to the proposal in~\cite{Johnson:2017asf}, by writing $r=r_++\zeta\sigma$ and $t= \tau/\zeta$, where, $Y(r=r_+)=0$ and $Y'(r=r_+) = 4\pi T_{\rm cr}$. The near horizon limit was obtained  in~\cite{Johnson:2017asf} by taking $\zeta \to 0$, while at the same time taking the large $q$ limit by holding  $\zeta q$ fixed. In the present case, in addition to the above limits, looking at the form of critical quantities in eqn. (\ref{eq:critical pt}) one also needs to take the limit $c_1 \rightarrow 0$ (or $m \rightarrow 0$ limit\footnote{$m \rightarrow 0$ or $c_0 \rightarrow 0$ limit may not be smooth, particularly, for the cases $k=0,-1$. For $k=1$ the results are not very different from~\cite{Johnson:2017asf}. Further, a large value of $m$ can also destabilize other thermodynamic quantities~\cite{PVMassIV}}, due to the requirement of $\epsilon >0$ for the existence of critical region) to get a consistent near horizon metric, for the case of general topology.  Thus, the metric in (\ref{eq:metric}) with the critical values inserted goes over to:
\begin{equation}
ds^2 = -{(4\pi {\widetilde T}_{\rm cr})\,\,\sigma}d\tau^2+\frac{1}{(4\pi {\widetilde T}_{\rm cr})}\frac{d\sigma^2}{\sigma }+d{\mathbb R}^{2}\, .
\end{equation} 
Here, ${\widetilde T}_{\rm cr}$ is $T_{\rm cr}$ in equation~(\ref{eq:critical pt}) with $q$ replaced by~$\tilde q = \zeta q$ and $c_1$ replaced by $\tilde c_1 = c_1/\zeta$, where both $\tilde q$ and $\tilde c_1$ are held fixed. Also, since  $l_{cr}$ diverges in this limit, the effective cosmological constant is zero  and $r_{\rm cr}$ also diverges in the above scaling limit, from eqn. (\ref{eq:critical pt}), the metric there is essentially flat. Thus, this new triple scaling limit results in a completely decoupled Rindler space-time with zero cosmological constant. We should mention here, that a large charge limit does not exist for the case when $c_1$ is negative, as the critical mass vanishes at the point noted in eqn. (\ref{qvanish}) and the critical temperature at this point takes the value $T_{cr}= \frac{c_0 \, c_1 m^2}{8 \pi }$, which is unphysical, due to the bound that needs to be satisfied by thermodynamic quantities~\cite{PVMassIV}. Thus, in the case of negative $c_1$, a suitable near horizon limit which gives a decoupled Rindler space as in~\cite{Johnson:2017asf}, does not exist in general.

\section{Conclusions} \label{conclusions}
The study of holographic heat engines pioneered in~\cite{Johnson:2014yja} has been an interesting area of research where thermodynamic properties of black holes, in particular, at the critical point of phase transition play a key role in various calculations of efficiency, including acting as a toy model for insight from statistical mechanics~\cite{Johnson:2017hxu}. In this context, taking certain parameters of the model black hole system to be large at the critical point, has been one of the ways to approach Carnot efficiency. There have been various attempts at using bench marking schemes to quantify the results of efficiency calculations of black hole heat engines.  One result which concerns this work, was carried out in~\cite{Hennigar:2017apu,Hendi:2017bys}, where it was shown that, for a given fixed cycle with $C_V = 0$, efficiency $\eta_k$ was shown to follow the order:  $ \eta_{\rm -1}^{\phantom{-1}} > \eta_{\rm 0}^{\phantom{0}} > \eta_{\rm +1}^{\phantom{+1}}$, which correspond to efficiencies of heat engines for black holes with hyperbolic, flat and spherical horizon topologies, respectively. With regard to topological heat engines in general, it was noted that  black holes with hyperbolic horizons have higher efficiency. The above results were actually noted at a generic point in the thermodynamic phase space of black holes. Our aim in this manuscript was to test whether the above topological order of efficiency of heat engines is still valid, when the black hole is on the verge of a critical point, where the efficiency is actually maximized. We showed that when the engine runs close to the critical point of a second order phase transition, and in particular, for case of massive coefficient $c_1 = 0$, efficiency $\eta$  follows the reverse order, i.e., $ \eta_{\rm -1}^{\phantom{-1}} < \eta_{\rm 0}^{\phantom{0}} < \eta_{\rm +1}^{\phantom{+1}}.$  
The reason for this is as follows. In the cases considered in~\cite{Hennigar:2017apu,Hendi:2017bys}, as per the chosen scheme, the highest and lowest pressure, as well as the volumes in the engine are fixed\footnote{though the variation with topology is captured by equation of state, the change is compensated by changes in temperature}, so that the work done is fixed,  while the heat inflow $Q_H$ changes with topology only via an overall contribution from the term $k \times (\text{positive quantity} )$. On the other hand, in the case of critical engines considered in this work,  the highest and lowest pressure, as well as volumes in the engine depend on topology in a non-trivial way, and contribute every term in $Q_H$ (see eq.~\ref{QH}), although, the work remains independent of topology. This explains the overall observed behavior of efficiency $\eta$ with topology $k$. Perhaps, we should mention here that although, we have chosen a specific rectangular thermodynamic cycle for our computations, using the arguments  in~\cite{Rosso:2018acz}, our results are actually valid for any other thermodynamic cycle obtained by infinitesimal deformations,  making our conclusions more general\footnote{We thank Felipe Rosso for useful communication in this regard}. \\

\noindent
Moreover, for the critical engines, the Carnot efficiency $\eta_{\rm C}^{\phantom{C}}$ and the ratio   $\frac{\eta}{\eta_{\rm C}^{\phantom{C}}}$ follow the order: $ \eta_{{\rm C}^{\phantom{C}}_{-1}} > \eta_{{\rm C}^{\phantom{C}}_{0}} > \eta_{{\rm C}^{\phantom{C}}_{+1}} $ and $\Big(\frac{\eta}{\eta_{\rm C}^{\phantom{C}}}\Big)_{\rm -1}^{\phantom{-1}} < \Big(\frac{\eta}{\eta_{\rm C}^{\phantom{C}}}\Big)_{\rm 0}^{\phantom{0}} < \Big(\frac{\eta}{\eta_{\rm C}^{\phantom{C}}}\Big)_{\rm +1}^{\phantom{+1}} $, which shows that the approach to Carnot efficiency is higher for the engine with  higher $k$. However, $\eta$ and $\eta_{\rm C}^{\phantom {C}}$  converge at large charge $q$ to $\frac{3}{19}$, a result which is independent of both the topology of horizon and massive gravity parameters, and, identical to the results noted for the case of charged black hole ~\cite{Johnson:2017hxu}.
Further, since $\eta$ is bounded by $\eta_{\rm C}^{\phantom{C}}$, and both depend on the graviton mass $m$, it is clear from the expressions of $\eta$ and $\eta_{\rm C}^{\phantom{C}}$ (see \ref{eq:work}, \ref{QH} and \ref{eq:itac}) and also from figure~\eqref{fig:meffect_k_plots} that when the massive coefficient $c_2$ is positive, one can take the $m \rightarrow \infty$ limit where both $\eta$ and $\eta_{\rm C}^{\phantom{C}}$ approach  the value $\frac{3}{19}$ (which is the same value obtained by taking the $q \rightarrow \infty$ limit), independent of horizon topology $k$.
 However, when the massive coefficient $c_2$ is negative, only the black holes with spherical topology ($k= +1$) show  the critical behavior provided $(1+ m^2 c_2 c_0^2)>0.$ In this case, $\eta_{\rm C}^{\phantom{C}} \rightarrow \infty$ when $(1+ m^2 c_2 c_0^2) \rightarrow 0$ (see figure~\eqref{fig:meffect_k_1_plots}). Hence, the requirement $\eta_{\rm C}^{\phantom{C}} \leq 1$, restricts the graviton mass $m$ to obey: $m^2 \leq \frac{1}{c_2 c_0^2} (\frac{L^2}{9 q^2 4^{1/3}} - 1)$. It is remarkable that efficiency of heat engines can be useful in obtaining a bound for the graviton mass. 
\vskip 0.4cm 

\noindent
Another point worth mentioning is that, from the equation of state~\eqref{eq of st}, the behavior of pressure $p$ with horizon topology $k$ is : $p_{\rm {+1}}^{\phantom{+1}} < p_{\rm {0}}^{\phantom{0}} < p_{\rm {-1}}^{\phantom{-1}} $, where as, the behavior of the critical pressure $p_{cr}$  with topology $k$ (see equation~\ref{eq:critical pt}) is : $p_{{\rm {cr}}^{\phantom{cr}}_{+1}} > p_{{\rm {cr}}^{\phantom{cr}}_0} > p_{{\rm {cr}}^{\phantom{cr}}_{-1}} $. This inverse behavior of pressure with topology $k$ still holds in the vicinity of the critical point of the second order phase transition. Since we kept our engine in the critical region and defined the operating pressures of the engine in terms of the critical pressure $p_{cr}$, the inverse order of the critical pressure $p_{cr}$ with topology $k$, is encoded in the topological order of efficiency $\eta$, and is the main reason for our engine to exhibit the inverse topological order of efficiency  in the critical region. If we move our engine to any other generic point away from the critical region in the phase space, then the usual topological order of efficiency follows i.e.,  $ \eta _{+1} < \eta _{0} < \eta _{-1}$. Since, pressure is inversely related to the AdS length scale $l$ (as $p=\frac{3}{8 \pi l^2}$), the interplay of cosmological constant and the horizon topology $k$ plays a crucial role in various features of heat engines in massive gravity theories. Thus, topological heat engines and their efficiency, either at the critical point or far from it can teach us much more about massive gravity theories, such as, giving bounds on graviton mass and the critical mass of the black hole (e.g., eqn. (\ref{Mcr})). 

\vskip 0.4cm 

\noindent
An interesting avenue to explore concerns the topological order of efficiencies in massive gravity from the point of view of AdS/CFT. From the point of view of holography, AdS length $l$ is related to the number of colors $N$ (rank of the gauge group) which indicates the available degrees of freedom of the theory. The precise connection between $l$ and $N$, depends on the family of the CFTs being considered~\cite{Maldacena:1997re,Johnson:2014yja,Kubiznak:2016qmn}. One would expect a non trivial interplay of the number of colors $N$ in the boundary CFTs corresponding to different topologies and its effect on the degrees of freedom. It would be nice to take these connections forward. There is another approach where AdS/CFT can give more insight in to the dual field theories. Very nice connections of entanglement entropy, renormalization group (RG) flow and holographic heat engines are proposed in~\cite{Johnson:2018amj}, where thermal journey in the PV-plane leads to holographic RG flow in CFT, corresponding to a tour in the space of field theories. Also, ratios of efficiency of heat engines in the bulk are related to the ratios of number of degrees of freedom in the boundary field theory, with interesting implications for entanglement entropy. These are interesting problems to pursue in future. This connection works for black holes with hyperbolic topology and it would be nice to explore extensions of the set up used in~\cite{Johnson:2018amj} for black holes with arbitrary topologies to make progress in our understanding through holography. The methods of holographic renormalization group developed for massive gravity theories in~\cite{Alishahiha:2010bw} may be helpful. 
 It is also tempting to study the efficiency of critical heat engines in other black hole systems considered in~\cite{Hennigar:2017apu}, such as, Kerr-AdS system etc., which support critical behavior for non-trivial topologies, and see if the dependence of efficiency on topology is valid in general. \\
 
 \vskip 0.2cm
 
\noindent
We also studied the critical region of black holes in massive gravity by analyzing the behavior of charged particles in the probe approximation, moving in the background of the critical hole. It was noted that 
there is an attractive potential binding the system together with no local minimum, when the mass  to charge ratio of the particle is equal to the critical mass to charge ratio of the black hole. We further showed that a fully decoupled Rindler space-time appears in the near horizon limit in a new triple scaling limit. However, as we also saw that such a near horizon limit may not always exist and is not universal, the implications of which might be addressed in holographic settings. 

\section*{Acknowledgements} 
We thank the referees for critical review and suggestions which improved the manuscript.

\section*{Appendix A}
\subsection*{Various scalings of the engine cycle :}
 The calculation of efficiency of heat engines when the working substance is on the verge of a second order phase transition is done by placing the thermodynamic cycle in the region close to critical point (which was chosen to be corner-3, as marked in 
figure-(\ref{fig:sample isotherms})). The next step is to  choose various scalings of the operating  pressures and volumes of our rectangular cycle, as done in eqn. (\ref{cyclecrit}). In this appendix, we provide three different scalings of critical quantities and show that the choice in eqn. (\ref{cyclecrit}) is rather special and other choices do not lead to desired set ups to understand approach of engine efficiency to Carnot efficiency. 


\begin{enumerate}
	\item 
	
	 \underline{General scaling of $p$ and $V$:}	\ 	Let us define the engine cycle as,
	\begin{eqnarray} \label{eq:general scaling}
	p_1 &= &p_2 = a p_{\rm cr}, \nonumber \\ 
	p_4 &= & p_3  =  p_{\rm cr},  \nonumber \\
	V_2 &=& V_3=V_{\rm cr},  \nonumber \\ \text{and} \, \, \, \,  V_1 &=& V_4= b V_{\rm cr} \ , 
	\end{eqnarray}
	where  the constants satisfy: $a >1$ and $0 < b <1$. Then, the work done is not a constant and depends on the parameters $( k, \ m, \ q )$ as:
	\begin{equation}
	W = \frac{q \sqrt{\epsilon}}{2\sqrt{6}}(a-1)(1-b)\, .
	\end{equation}	
The efficiency $\eta$ and the Carnot efficiency $\eta_{\rm C}^{\phantom{C}}$ are given respectively as
	\begin{eqnarray}
	\eta &=& \frac{(a-1)(1 + b^{1/3} + b^{2/3}) b^{1/3}}{(6+a)b^{1/3} + ab^{2/3} +ab-1}, \\
	\eta_{\rm C}^{\phantom{C}} & = & \frac{1-6b^{2/3} + 5b + 3ab-3b^{4/3}}{5b+3ab},  
	\end{eqnarray}
on the other hand are independent of the parameters $( k, \ m, \ q )$. Hence the choice of critical quantities  of the engine in eqn. (\ref{eq:general scaling}) is not a useful scaling to study the topological effects on efficiency $\eta$.
\\

\item 

 \underline{Scaling of pressure:}		\ 	Let us define the engine cycle as,
	\begin{eqnarray}
	p_1 &= &p_2 = \Big(1+\frac{L}{q\sqrt{\epsilon}}\Big) p_{\rm cr}, \nonumber \\ 
	p_4 &= & p_3  =  p_{\rm cr},  \nonumber \\
	V_2 &=& V_3=V_{\rm cr},  \nonumber \\ \text{and} \, \, \, \,  V_1 &=& V_4= b V_{\rm cr} \ , 
	\end{eqnarray}
	where, $ \frac{L}{q\sqrt{\epsilon}} > 0$ and $0 < b <1$. This scaling again gives the  work done $ W $ to be a constant as
	\begin{equation}
	W = \frac{L (1-b)}{2\sqrt{6}},
	\end{equation}	
though depending on two parameters. Further, as shown in figures~\eqref{fig:psacling_q plots}  and~\eqref{fig:psacling_m plots},  the  efficiency $\eta$ follows the topological order i.e., $\eta_{+1} < \eta_{0} < \eta_{-1}$, which is similar to the order followed at a generic point (not at critical point)~\cite{Hennigar:2017apu,Hendi:2017bys}.
\begin{figure}[h!]
	{\centering
		\subfloat[]{\includegraphics[width=2.5in]{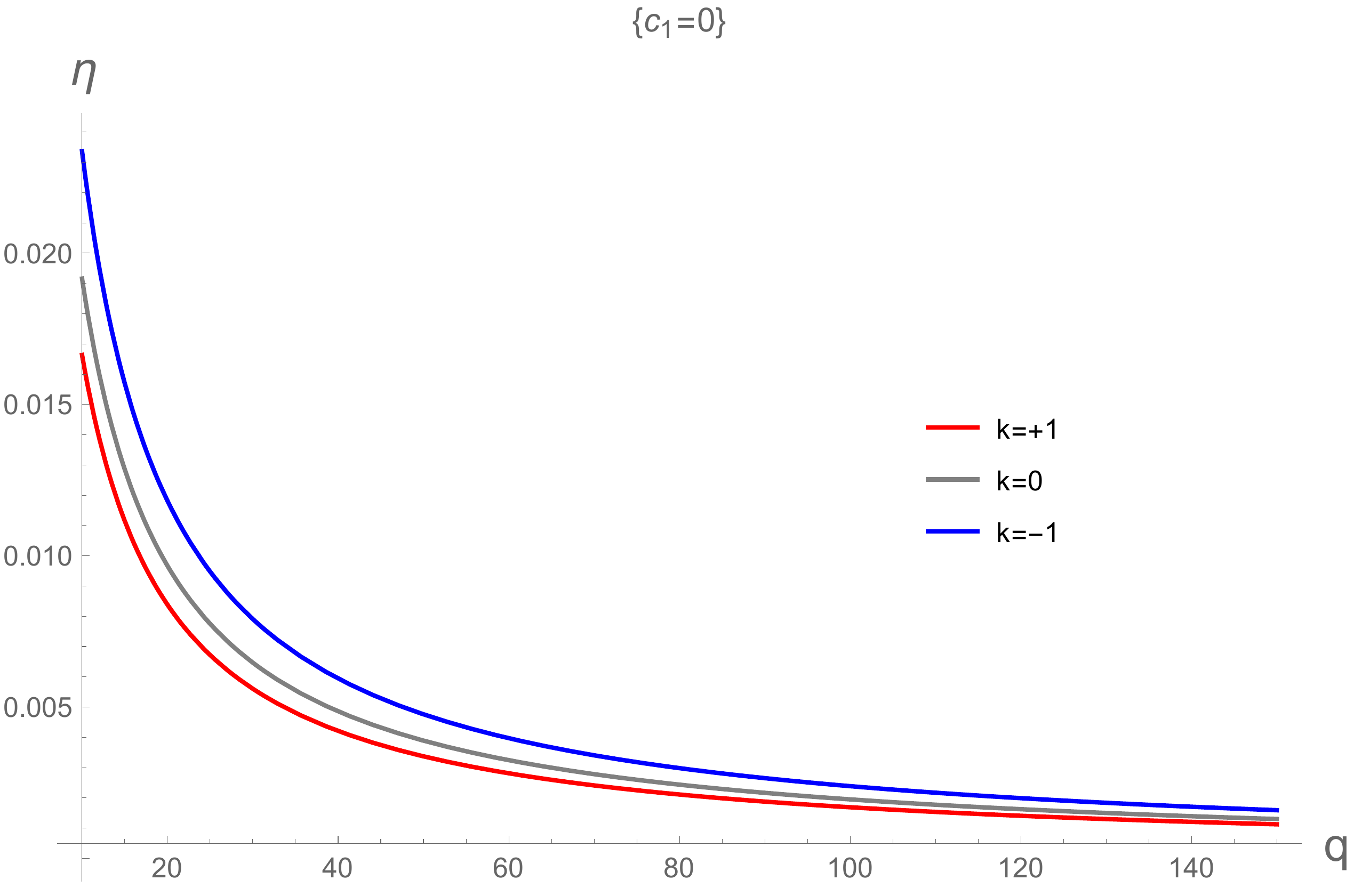}}\hspace{0.75cm}	
		\subfloat[]{\includegraphics[width=2.5in]{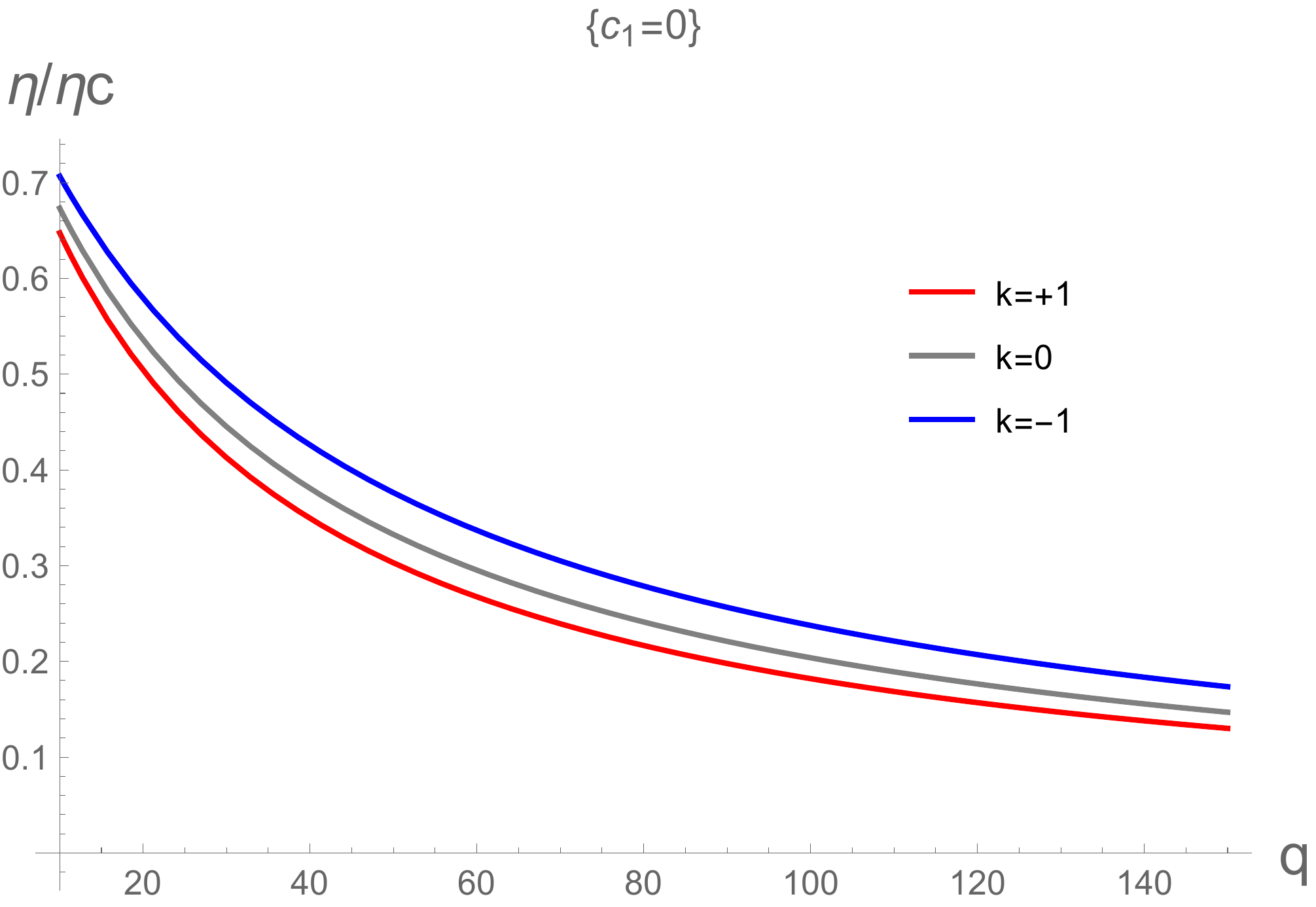}}				
		\caption{\footnotesize In the case of massive coefficient $c_1=0$, the effects of topology $k$ and charge $q$ on (a) $\eta$,  and (b) $\eta/\eta_{\rm C}$. (Here, the parameters $L=m=c_0=1$, $b =1/2$ and  $c_2 =3$, are used.) 
		}   \label{fig:psacling_q plots}
	}
\end{figure}
\begin{figure}[h!]
	{\centering
		\subfloat[]{\includegraphics[width=2.5in]{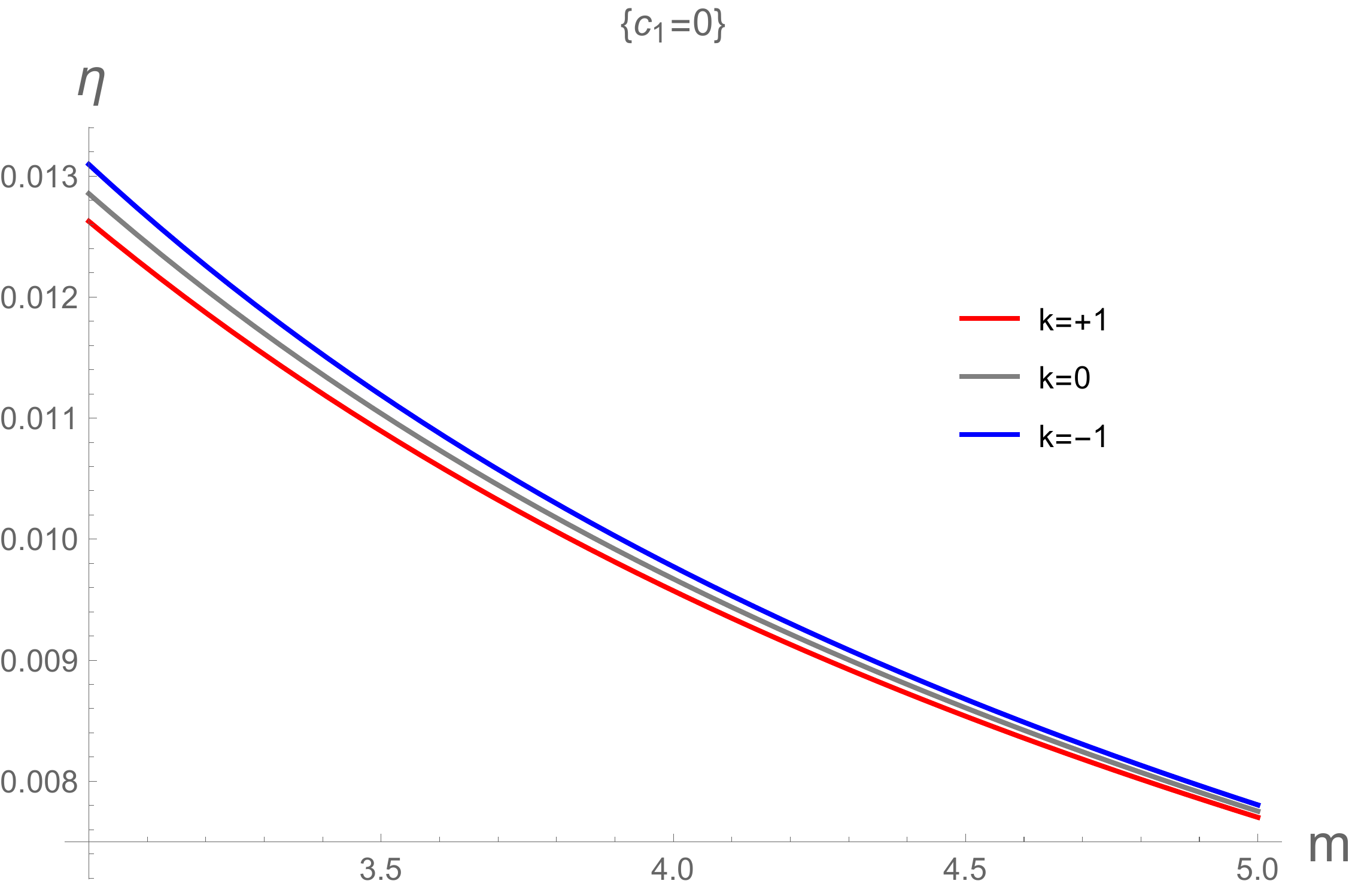}}\hspace{0.75cm}	
		\subfloat[]{\includegraphics[width=2.5in]{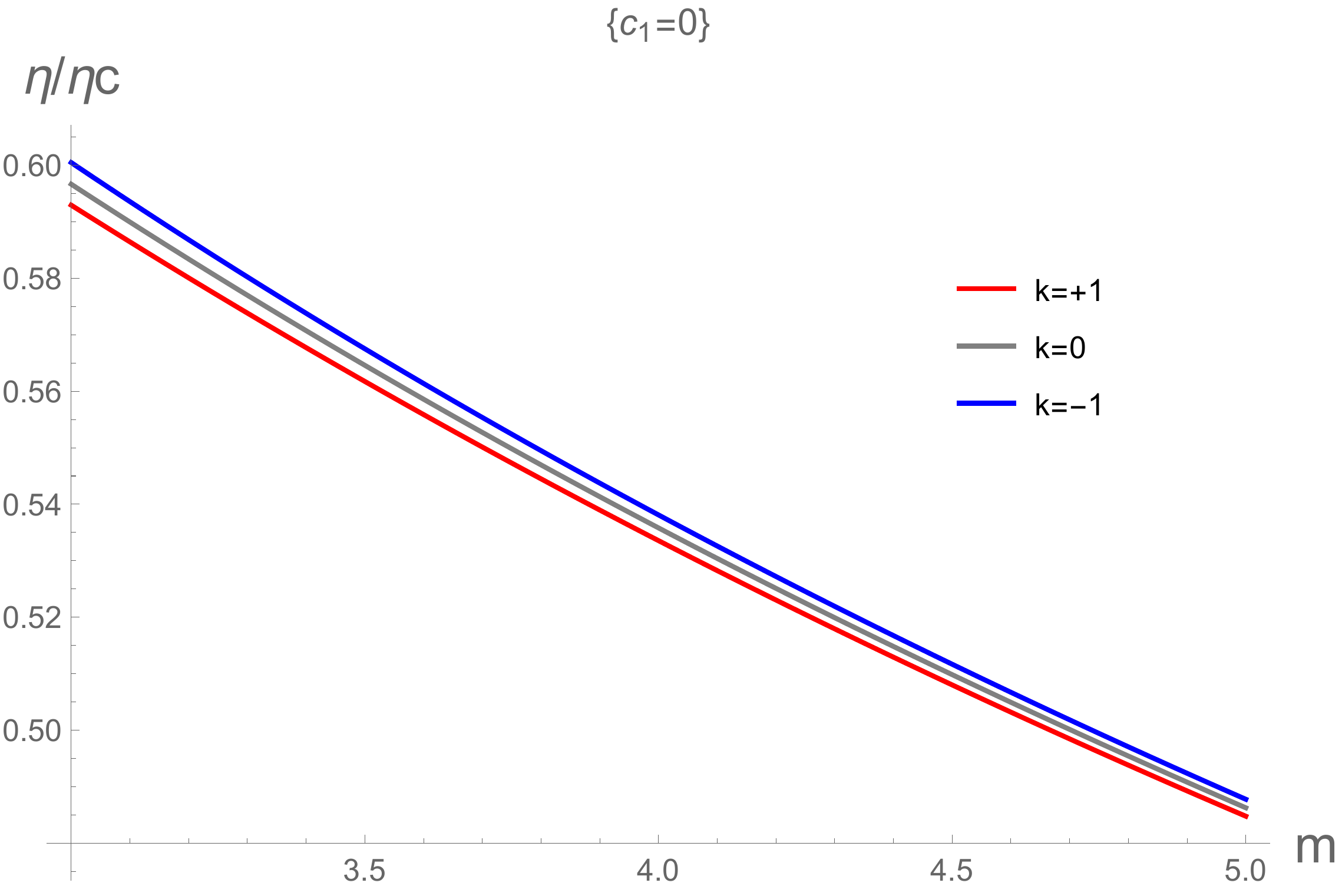}}				
		\caption{\footnotesize In the case of massive coefficient $c_1=0$, the effects of topology $k$ and the graviton mass $m$ on (a) $\eta$,  and (b) $\eta/\eta_{\rm C}$. (Here, the parameters $L=c_0=1$, $b =1/2$, $q=5$ and  $c_2 =3$, are used.) 
		}   \label{fig:psacling_m plots}
	}
\end{figure}
However, this scaling of the engine leads to an overall decrease in the efficiency $\eta$, with an undesirable feature that the ratio $\frac{\eta}{\eta_{\rm C}^{\phantom{C}}}$ never approaches unity, in the limit of large parameters, such as, large $q$ or $m$. Thus,
this scaling is again not useful to consider while studying approach to Carnot efficiency. 	

\item 

 \underline{ Scaling of $p$ and $V$:}\ 	Let us now define the engine cycle as,
\begin{eqnarray}
p_1 &= &p_2 = \Big(1+\frac{L}{q\sqrt{\epsilon}}\Big) p_{\rm cr}, \nonumber \\ 
p_4 &= & p_3  =  p_{\rm cr},  \nonumber \\
V_2 &=& V_3=V_{\rm cr},  \nonumber \\ \text{and} \, \, \, \,  V_1 &=& V_4= \Big(1-\frac{L}{q\sqrt{\epsilon}}\Big) V_{\rm cr} \ , 
\end{eqnarray}
where, $  0 < \frac{L}{q\sqrt{\epsilon}} <1 $. This scaling gives the  work done $ W $, which is not constant and depends on the parameters $(k, \ q, \ m)$, given by 
\begin{equation}
W = \frac{L^2}{2 q \sqrt{6 \epsilon}}\, .
\end{equation}	
As shown in figures~\eqref{fig:pvsacling_q plots}  and~\eqref{fig:pvsacling_m plots},  the  efficiency $\eta$ again follows the  topological order i.e., $\eta_{+1} < \eta_{0} < \eta_{-1}$(which is similar to the result in~\cite{Hennigar:2017apu,Hendi:2017bys} at a generic point in thermal journey).
\begin{figure}[h!]
	{\centering
		\subfloat[]{\includegraphics[width=2.5in]{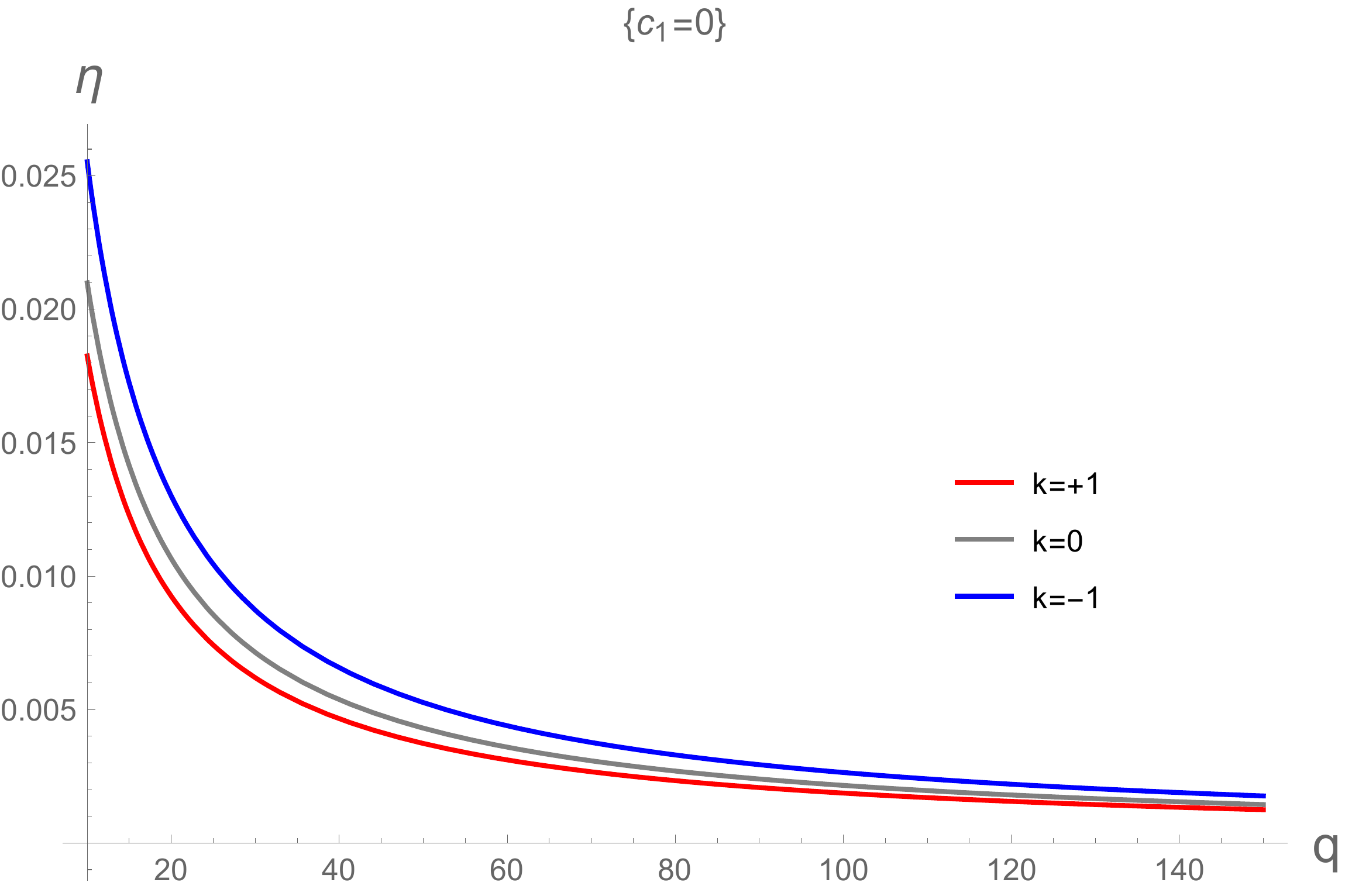}}\hspace{0.75cm}	
		\subfloat[]{\includegraphics[width=2.5in]{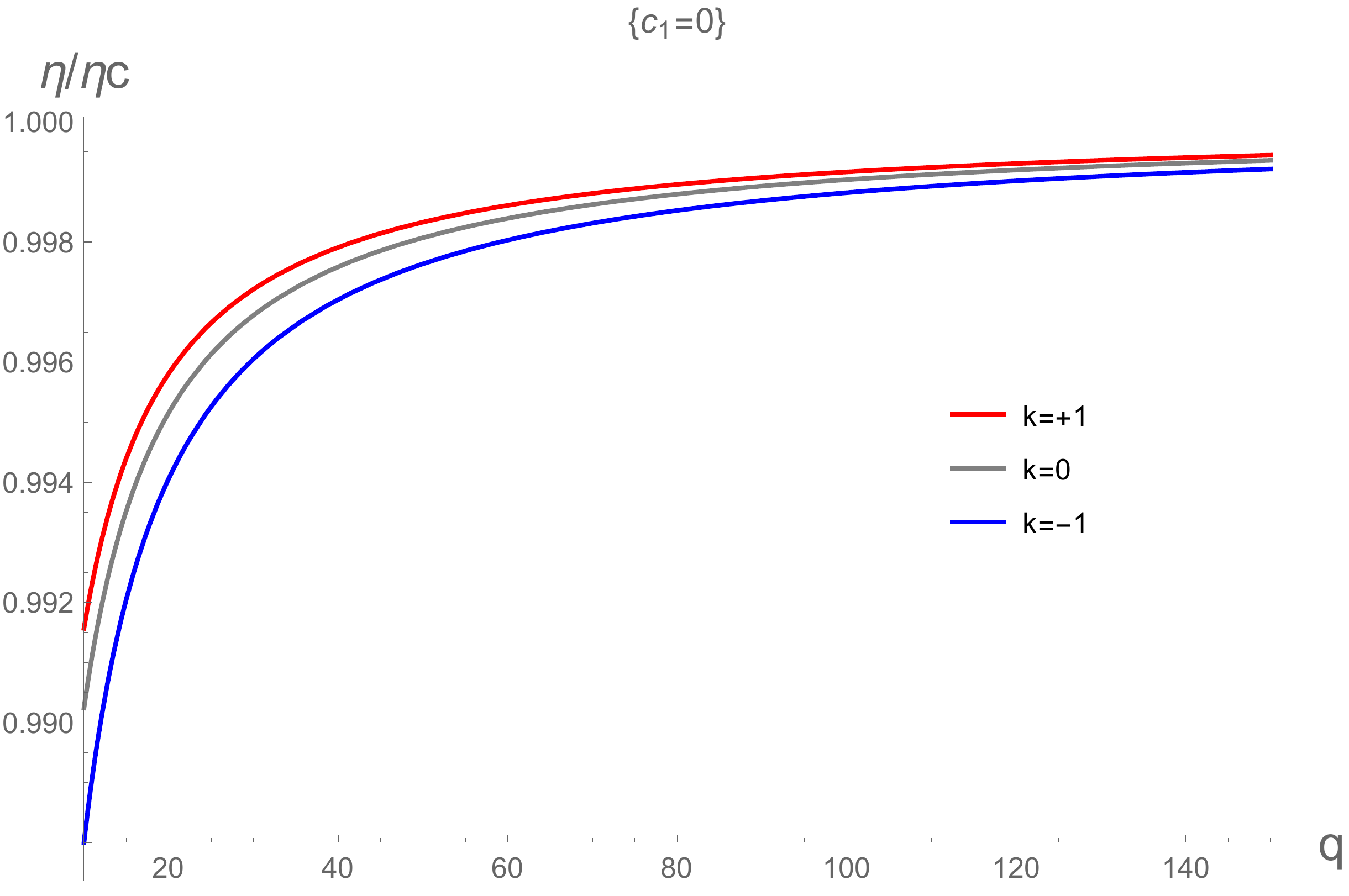}}				
		\caption{\footnotesize In the case of massive coefficient $c_1=0$, the effects of topology $k$ and charge $q$ on (a) $\eta$,  and (b) $\eta/\eta_{\rm C}$. (Here, the parameters $L=m=c_0=1$,  and  $c_2 =3$, are used.) 
		}   \label{fig:pvsacling_q plots}
	}
\end{figure}
\begin{figure}[h!]
	{\centering
		\subfloat[]{\includegraphics[width=2.5in]{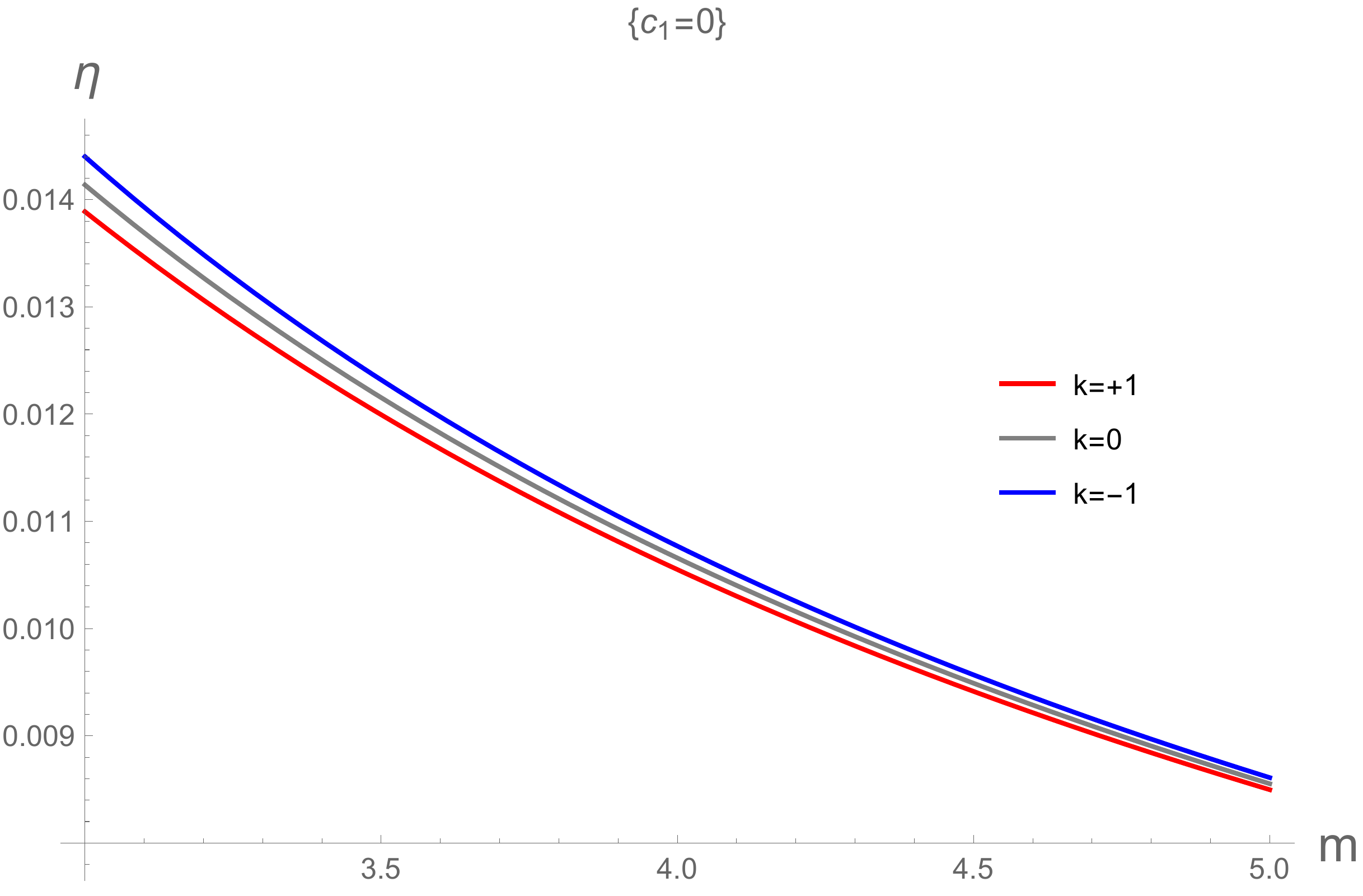}}\hspace{0.75cm}	
		\subfloat[]{\includegraphics[width=2.5in]{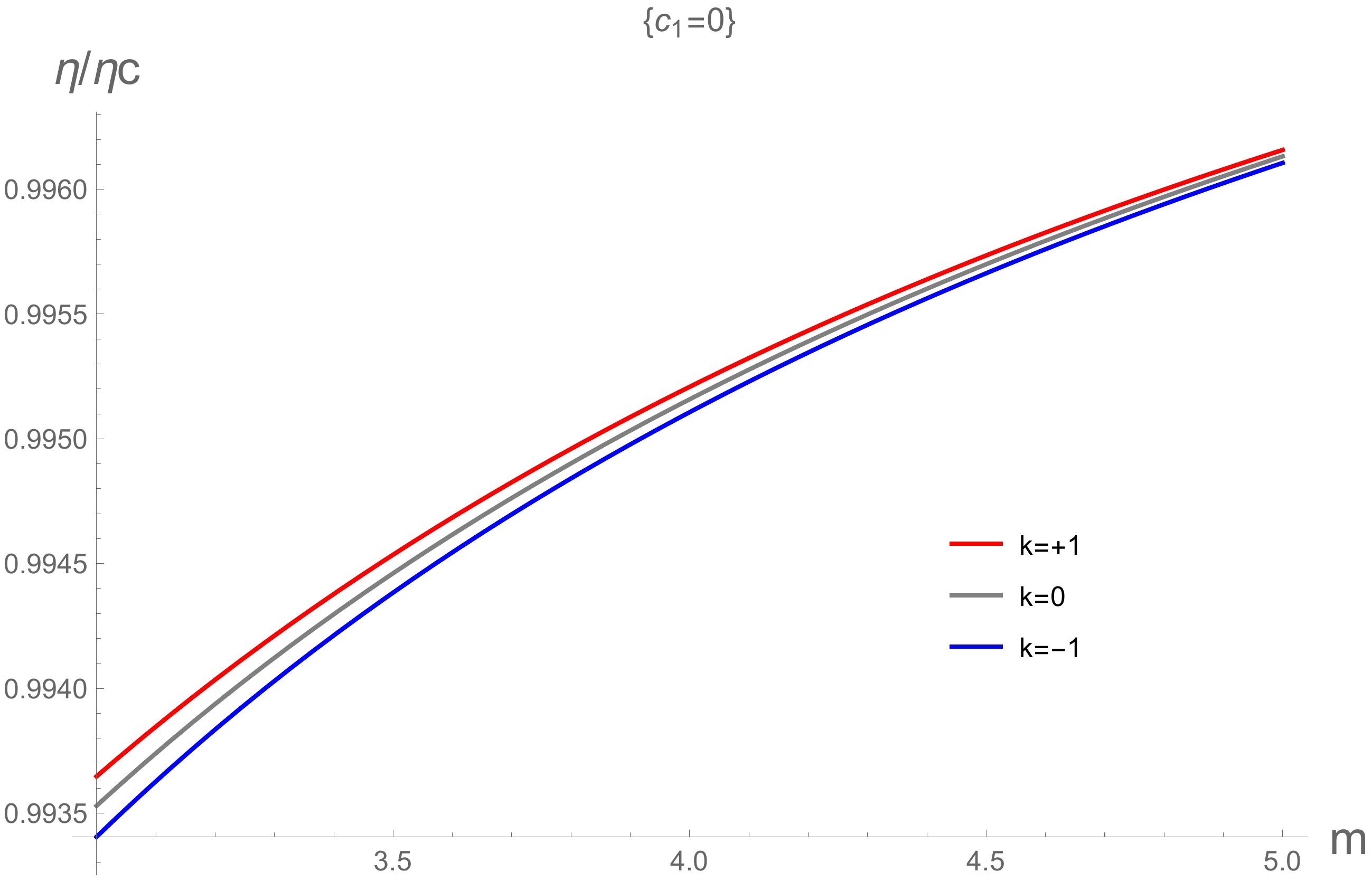}}				
		\caption{\footnotesize In the case of massive coefficient $c_1=0$, the effects of topology $k$ and the graviton mass $m$ on (a) $\eta$,  and (b) $\eta/\eta_{\rm C}$. (Here, the parameters $L=c_0=1$,  $q=5$ and  $c_2 =3$, are used.) 
		}   \label{fig:pvsacling_m plots}
	}
\end{figure}
This scaling of the engine again leads to decrease in the efficiency $\eta$, through the ratio $\frac{\eta}{\eta_{\rm C}^{\phantom{C}}} \rightarrow 1$. However, the work done $W \rightarrow 0$, as the charge or the graviton mass is taken to be large. Hence, this scaling again is not a useful to consider.
\end{enumerate} 

\noindent
 Therefore, these scalings of the operating pressures and volumes of the engine cycle have their own limitations and the particular scaling that we used in equation~\eqref{cyclecrit} is a nice choice envisaged in~\cite{Johnson:2017hxu} for approaching the Carnot efficiency in the large parameter limit.

\section*{Appendix B}
	\subsection*{Scheme independent behavior of efficiency $\eta$ :} Before computing the efficiency $\eta$ of our rectangular engine cycle in $p-V$ plane, we must decide on the parameters of the cycle we specify and hold fixed. There are several different choices called as schemes. Equation~\eqref{cyclecrit} represents one such scheme (among others) in which we specify and hold fixed the operating pressures $(p_1, \ p_3)$, and volumes $(V_1, \ V_3)$. We can choose any other scheme for our rectangular cycle, however, the behavior of the efficiency $\eta$ with the parameters of theory may be scheme dependent (see, for example, \cite{Johnson:2015ekr}). However, to get a scheme independent behavior  of efficiency $\eta$ for our engine, we must choose the appropriate scaling to the specified parameters of the cycle. This can be done, for example, as follows.
	\vskip 0.3cm 
	\noindent  Let us illustrate this by choosing a new scheme in which  we specify and hold fixed the operating pressures $(p_1, \ p_4)$ and temperatures $(T_3, \ T_4)$ of the cycle, as:
	\begin{eqnarray}\label{eq:scheme scaling}
	p_1 &= &p_2 = a p_{\rm cr}, \nonumber \\ 
	p_4 &= & p_3  =  p_{\rm cr},  \nonumber \\
	T_3 &=& T_{\rm cr},  \nonumber \\ \text{and} \, \, \, \,  T_4 &=&  b T_{\rm cr} \ , 
	\end{eqnarray} where the constants, $a>1$ and $0<b<1$. Then, we can use the equation of state~\eqref{eq of st} to compute the corresponding volumes of the cycle, which are:
	\begin{equation*}
	V_3 = V_2 = V_{cr},
	\end{equation*} while, the expression for $V_4(= V_1)$ is a complicated function of $b$ (for example, for $b=1/2$, $V_4 = (\frac{1.11813}{\sqrt{6}})^3 V_{cr}$). Thus, this scaling in equation~\eqref{eq:scheme scaling} is  equivalent to the general scaling given in the equation~\eqref{eq:general scaling}, and gives the $\eta$ and $\eta_{\rm C}^{\phantom{C}}$, which are  independent of the parameters $(k, \ m, \ q)$. However, if we fix $a = 3/2$, and $b=\frac{1}{8}\Big( 6(1-\frac{L}{q\sqrt{\epsilon}})^{-1/3} -(1-\frac{L}{q\sqrt{\epsilon}})^{-1} + 3(1-\frac{L}{q\sqrt{\epsilon}})^{1/3}\Big)$, then $V_3 = V_{cr}$ and $V_4 =(1-\frac{L}{q\sqrt{\epsilon}}) V_{cr}$, which exactly reproduces the previous scheme scalings given in equation~\eqref{cyclecrit}. In addition, the inverse topological order of efficiency found in this work in eqn. (\ref{etaorder}) is generated again, with in this new scheme.
	\vskip 0.3cm
	 \noindent Therefore, we can choose any scheme in which we specify and hold some parameters of the engine cycle fixed, where the equation of state can readily be used to compute the other parameters of the cycle. If we choose the appropriate scaling to the specified parameters  of any scheme, then, we can get universal results on the behavior of efficiency of the critical engine, which is guaranteed to be scheme independent.  
\bibliographystyle{apsrev4-1}
\bibliography{massive}
\end{document}